\documentclass[aps,showpacs,twocolumn,amsmath,amssymb,amsfonts,eqsecnum]{
revtex4-1}
\usepackage{graphicx}
\usepackage{color}
\usepackage{natbib}
\usepackage{epstopdf}
\usepackage{hyperref}

\begin{document}
\title{Casimir force in the rotor model with twisted boundary conditions}
\author{Jonathan Bergknoff$^{1}$\thanks{e-mail: jbergk@physics.ucla.edu}, Daniel Dantchev$^{1,2}$\thanks{e-mail:
daniel@imbm.bas.bg} and  Joseph Rudnick$^{1}$\thanks{e-mail:
jrudnick@physics.ucla.edu}} \affiliation{ $^1$ Department of Physics and Astronomy, UCLA, Los Angeles,
California 90095-1547, USA,\\$^2$Institute of
Mechanics - BAS, Academic Georgy Bonchev St. building 4,
1113 Sofia, Bulgaria
}
\date{\today}

\begin{abstract}
We investigate the three dimensional lattice $XY$ model with nearest neighbor interaction.
%for which the dynamical variables are two dimensional vectors able
%to rotate only in a plane.
The vector order parameter of this system lies on the vertices of a cubic lattice, which is embedded in a system with a film geometry. The  orientations of the vectors are fixed at the two opposite sides of the film. The angle between the vectors at the two boundaries is $\alpha$ where $0 \le  \alpha \le \pi$. We make use of the mean field approximation to study the mean length and orientation of the vector order parameter throughout the film---and the Casimir force it generates---as a function of the temperature $T$, the angle $\alpha$, and the thickness $L$ of the system. Among the results of that calculation are a Casimir force that depends in a continuous way on both the parameter $\alpha$ and the temperature and that can be attractive or repulsive. In particular, by varying $\alpha$ and/or $T$ one controls \underline{both} the sign \underline{and} the magnitude of the Casimir force in a reversible way. Furthermore, for the case $\alpha=\pi$, we discover an additional phase transition occurring only in the finite system associated with the variation of the orientations of the vectors.

\end{abstract}

\maketitle

\section{Introduction}

For an $O(n),n\geq 1$ model of a $d$-dimensional system with a temperature $T$ and geometry $\infty^{d-1}\times L$ the thermodynamic Casimir force is defined by  \cite{E90book}, \cite{BDT2000}
\begin{equation}
F_{\rm Casimir}^{(\tau )}(T,L)=-\frac{\partial f_{\rm ex}^{(\tau )}(T,L)}{\partial L}%
\text{,}  \label{def}
\end{equation}
where $f_{\rm ex}^{(\tau)}(T,L)$ is the excess free energy
\begin{equation}
f_{\rm ex}^{(\tau )}(T,L)=f^{(\tau)}(T,L)-Lf_b(T)\text{,}  \label{fexd}
\end{equation}
and the superscript $\tau $ denotes the boundary conditions. Here $f^{(\tau )}(T,L)$ is the full free energy per unit area of such a system subjected to the boundary conditions $\tau $ and $f_b
$ is the bulk free energy density. Accumulated evidence
\cite{E90book,K94,BDT2000,KD92a,D98,BE95,K97,HSED98,K99,GC99,GC2002,ML2001,
UBMCR2003,HK92,ES94,
DKD2003,SHD2003,DK2004,ZRK2004,FYP2005,DDG2006,DGS2006,GSGC2006,ZSRKC2007,
DSD2007,VGMD2007,MGD2007,H2007,H2009,GD2008,HHGDB2008,B2007,KK2006,DG2009,
GMHNHBD2009,MMD2010,
AM2010,H2010,H2011} supports the conclusion that if the boundary conditions are
identical---or sufficiently similar---at both surfaces bounding the system,
$F_{\rm Casimir}^{(\tau )}$ will be negative. In the case of a fluid confined
between identical walls this implies that the net force between those walls due
to the fluid will be {\em attractive} for large separations. On the other hand,
if the fluid wets one of the walls while the other wall prefers the vapor phase,
then the Casimir force is repulsive. This implies that if the boundary
conditions differ sufficiently the Casimir force can be expected to be positive,
or \emph{repulsive}, for the entire range of  thermodynamic parameters. In the
intermediate case in which one of the surfaces of a system belonging to the
Ising universality class has a weak preference for one of the phases of the
fluid while the other one exhibits a strong preference, or for a given ratio
of the surface fields and/or of surface enhancements on both
surfaces, it has been recently demonstrated \cite{AM2010,MMD2010,H2011} that one
can observe much richer behavior, with the Casimir force changing its sign once,
or even twice \cite{MMD2010}, as the temperature is adjusted. In addition, in
\cite{HGS2011} it has been
shown via Monte Carlo simulations that in a system with a geometry
$L_\perp \times L_{\|}^2$ subject to periodic boundary conditions both the
magnitude and the sign of the Casimir force depend on the aspect ratio
$a_r=L_{\perp}/L_{\|}$. In this case general arguments have been advanced to  suggest that
at the bulk critical point the Casimir force vanishes for $a_r=1$
and becomes repulsive for $a_r>1$. These results are supported by exact
calculations for the two-dimensional Ising model. For further information regarding the
results currently available on the critical Casimir effect the interested reader is
referred to general reviews \cite{K94,BDT2000} as well as articles devoted to
specific aspects of the critical Casimir force \cite{K99,G2009,TD2010rev,GD2011}.

The critical Casimir effect and the corresponding Casimir force discussed
above are due to spatial restrictions imposed on the thermal average of the order parameter and on its long ranged fluctuations in a system undergoing a second order phase transition.
Based on analogy, some hints regarding the general behavior of the thermodynamic
Casimir force can also be extracted  from the information available on the
quantum Casimir effect \cite{C48} which is due to the spatial restrictions
imposed on the possible fluctuations of the electromagnetic field. A series of
reviews devoted to different aspect of this latter effect are available
\cite{MT97,M94,KG99,BMM2001,M2001,M2004,L2005,GLR2008,BKMM2009,KMM2011,
RCJ2011}. For the quantum Casimir effect it has been demonstrated that if the
boundary conditions $\tau$ are symmetric so that reflection positivity holds,
the Casimir force $F_{\rm Casimir}^{(\tau )}$ is attractive \cite{B2007,KK2006}.
According to the theory of Dzyaloshinskii, Lifshitz, and Pitaevskii \cite{DLP61}
in any system with a slab-like structure in which a material $B$ separates two
identical half-spaces $A\equiv C$ the force is attractive as well. When $B$ is a
vacuum, again according to \cite{DLP61}, this remains true even when the
half-spaces $A$ and $C$ are not identical. This prediction, up to now, has been
verified for all materials for which the Casimir force has been measured.
Theoretical predictions exist, however, suggesting that in the latter case a
repulsive Casimir force can be generated by special selection of the material
properties of  $A$ and $C$; see, e.g., \cite{KKMR2002}. However, such a
situation has not been experimentally realized. The omnipresence of attractive
quantum Casimir electromagnetic force for objects in vacuum or air affects the
work of micro and nano-machines \cite{BR2001,L2005,GLR2008,RCJ2011} and might
cause sticking of their working surfaces. The possibility of realizing and controlling
a repulsive critical Casimir force might be one of the ways of overcoming the
above-mentioned difficulties.

In an attempt to shed additional light on the influence of differing boundary conditions on
the critical Casimir force, we consider a film system with $\infty^{d-1}\times
L$ geometry  consisting of local dynamical variables, say magnetic moments,
possessing $O(2)$ symmetry and constrained to lie in the $x$-$y$ plane. The
moments in one of the bounding surfaces are constrained to point in the same
direction in that plane and to be oriented at an angle  $\alpha$ with respect to
the similarly aligned spins in the other bounding surface. Furthermore, in the
case in which the moments have variable amplitudes, those amplitudes are fixed
at a non-zero value. Alternatively, one might think of the studied system as
a lattice gas of elongated, say, rod like molecules embedded on a lattice.
We investigate both the equilibrium behavior of those moments (or
molecules)---and the Casimir force that arises as a result of that behavior---as
a function of $T$ and $\alpha$. As we will see, among the results of our
calculation are a Casimir force that depends in a continuous way on both the
parameter $\alpha$ and the temperature and that can be attractive or repulsive.
In particular, by varying $\alpha$ and/or $T$ one controls both the sign and the
magnitude of the Casimir force in a reversible way.

We will refer to the boundary conditions described above as ``twisted'' boundary
conditions.   Subject to them, the moments within the system settle into a state
in which they rotate with respect to each other as the region between the
boundaries is traversed, creating a diffuse interface within it. The normalized
excess free energy per unit area of the system, $f_{\rm ex}^{(\alpha)}$, can be
related to the corresponding quantity in a system with zero twist of the
moments, which we will term a system with $(+,+)$ boundary conditions. The
quantified version of this relationship is
\begin{equation}
\label{falpha}
f^{(\alpha)}_{\rm ex}(T,L)=f^{(+,+)}_{\rm ex}(T,L)+\frac{1}{2L}\;\alpha^2 \,\Upsilon^{(\alpha)}(T,L),
\end{equation}
where $\Upsilon^{(\alpha)}(T,L)$ is the finite-size helicity modulus \cite{DG2009,D93,FBJ73} that characterizes the energy of the  system related to the diffuse interface in it. The excess free energy $f^{(\alpha)}_{\rm ex}(T,L)$ can also be resolved into regular and singular parts:
\begin{equation}
\label{fexdecomp}
f^{(\alpha)}_{\rm ex}(T,L)=f^{(\alpha)}_{\rm ex, \;reg}(T,L)+f^{(\alpha)}_{\rm ex,\;sing}(T,L).
\end{equation}
In the case of the singular part of the excess free energy in the vicinity of the bulk critical point one has,  according to  finite-size scaling theory \cite{BDT2000},
\begin{equation}\label{fexsing}
f_{\rm ex, \;sing}^{(\alpha)}(T,L)=L^{-(d-1)}X_{\rm ex}^{(\alpha)}\left(x_t\right)
\end{equation}
As a consequence, neglecting the ``background''  contribution to the Casimir force, one obtains
\begin{equation}\label{CasGeneral}
F_{\rm Casimir}^{(\alpha)}(T,L)=L^{-d}X_{\rm Cas}^{(\alpha)}\left(x_t\right).
\end{equation}
Here $x_t=a_t tL^{1/\nu }$ is the temperature scaling variable, $t=(T-T_c)/T_c$
is the reduced temperature, $a_t$ is a nonuniversal scaling factor, while
$X_{\rm ex}^{\left( \alpha\right) }$  and $X_{\rm Cas}^{\left( \alpha\right) }$
are {\em universal} (albeit geometry-dependent) scaling functions and $\nu $ is
the corresponding (universal) scaling exponent that characterizes the
temperature divergence of the bulk two-point correlation length, $\xi$, when one
approaches the bulk critical temperature from above, i.e. $\xi(t\rightarrow
0^+)\simeq \xi_0^+ t^{-\nu}$ with $\xi_0^+$ being some system dependent metric
factor. For the behavior of $\Upsilon^{(\alpha)}(T,L)$ near $T_c$ from (\ref{falpha}) and
(\ref{fexsing}) one derives
\begin{equation}\label{helmodfss}
\beta \Upsilon^{(\alpha)}(T,L)  =L^{-(d-2)}X_\Upsilon^{(\alpha)}\left(x_t\right),
\end{equation}
where $X_\Upsilon^{(\alpha)}$ again is universal scaling function. Requiring a
$L$-independent behavior of $\Upsilon$ in the limit $L\to \infty$, one obtains
$\Upsilon(T)\equiv \lim_{L\to\infty}\Upsilon^{(\alpha)}(T,L)$, with $\Upsilon(T)\ge 0$,
$X_\Upsilon^{(\alpha)}\left(x_t\right)\sim |x_t|^{(d-2)\nu}$ and, thus, $\Upsilon(t)\sim
|t|^{(d-2)\nu}$, which is in a complete agreement with \cite{FBJ73}.

 When such a diffuse interface is present within the system and $T<T_c$ from Eq. (\ref{falpha}) it is easy to see that
\begin{equation}
F_{\rm Casimir}^{\left(\alpha\right) }(T< T_c)\simeq  \frac{1}{2} \alpha^{2} \Upsilon(T) L^{-2},\; L\to\infty. \label{fssclowTas}
\end{equation}
Since $\Upsilon(T)\ge 0$, Eq. (\ref{fssclowTas}) implies that the Casimir force will be {\it repulsive} and, for $d>2$, much {\it stronger}, of the order of $L^{-2}$, than in systems with a compact interface where it is either of the order of $L^{-d}$, or smaller.

The structure of the article is as follows. In the next section \ref{latticemodel} we define a lattice three-dimensional mean-field XY model and present numerical results for the behavior of the Casimir force within it. Section \ref{GLMFmodel} presents analytical results for the scaling function of the Casimir force within the Ginzburg-Landau mean-field theory of the three-dimensional XY model. In both sections \ref{latticemodel} and \ref{GLMFmodel} we find interesting behavior of the force at a temperature $T_{\rm kink}$ below the critical one of the bulk system when $\alpha$ approaches $\pi$. We study this special case in section \ref{alphaequalpicase}. We deduce the existence of an additional second-order phase transition that is specific to this finite system. The article closes with a discussion presented in section \ref{discussion}. Technical details of the derivations are presented in appendixes at the end of the article.

\section{The Casimir force in the lattice three-dimensional mean-field XY model}
\label{latticemodel}

Consider a lattice of dimensions $\infty^{d-1}\times L$, with each site
populated by an $O(2)$ fixed-length magnetic moment of magnitude $m$. We split
up the lattice into $(d-1)$-dimensional planes labeled ${1,\ldots,N}$, where
$L=N a$, with $a$ being the lattice constant taken in the remainder to be equal
to one. By translational invariance and neglecting the fluctuation within the
planes, all moments in plane $i$ must take the same value, equal to their mean
value, and must point in the same direction. However, due to the anisotropy
along the finite dimension, the moments will vary between planes. Let the moment
in plane $i$ be ${\bf m}_i$. We take a nearest-neighbor coupling with strength
$J$ both in the plane and out of it, and so the energy of a moment in plane $i$
will be
\begin{equation}
U_i=-J{\bf m}_i\cdot(2(d-1){\bf m}_i+{\bf m}_{i-1}+{\bf m}_{i+1})
\end{equation}
or, defining an effective magnetic field ${\bf H}_i=J(2(d-1){\bf m}_i+{\bf m}_{i-1}+{\bf m}_{i+1})$ at that site, $U_i=-{\bf m}_i\cdot{\bf H}_i$.

Approximating the moment ${\bf m}_i$ as being isolated, in an external magnetic field ${\bf H}_i$, we can assign it the local partition function
\begin{equation}
Z_i=\int_0^{2\pi} d\theta\ e^{\beta mH_i\cos\theta}=2\pi I_0(\beta mH_i)
\end{equation}
with $\beta=(k_B T)^{-1}$ and $\theta$ the angle between ${\bf H}_i$ and ${\bf m}_i$. On average, the component of ${\bf m}_i$ along ${\bf H}_i$ is
\begin{equation}
\langle m\cos\theta\rangle=\frac{1}{\beta}\frac{d\ln Z}{dH_i}=m\frac{I_1(\beta
mH_i)}{I_0(\beta mH_i)},
\end{equation}
where $I_0$ and $I_1$ are  the corresponding modified Bessel functions of the
first kind,
while the component normal to ${\bf H}_i$ is
\begin{equation}
\langle m\sin\theta\rangle=\frac{1}{Z}\int_0^{2\pi} d\theta\ m\sin\theta\ e^{\beta mH_i\cos\theta}=0
\end{equation}
so that the averaged moment is entirely along ${\bf H}_i$. Inserting the definition of ${\bf H}_i$ in terms of neighboring moments, we see that ${\bf m}_i$ must satisfy the equation
\begin{multline}
{\bf m}_i = \frac{2(d-1){\bf m}_i+{\bf m}_{i-1}+{\bf m}_{i+1}}{|2(d-1){\bf m}_i+{\bf m}_{i-1}+{\bf m}_{i+1}|}\\
\times R\left(\beta mJ\left|2(d-1){\bf m}_i+{\bf m}_{i-1}+{\bf m}_{i+1}\right|\right)m \label{latticeeom}
\end{multline}
in the mean-field approximation (all quantities now implicitly averaged), with $R(u)=I_1(u)/I_0(u)$.

\begin{figure*}[ht]
\includegraphics[angle=0,width=\columnwidth]{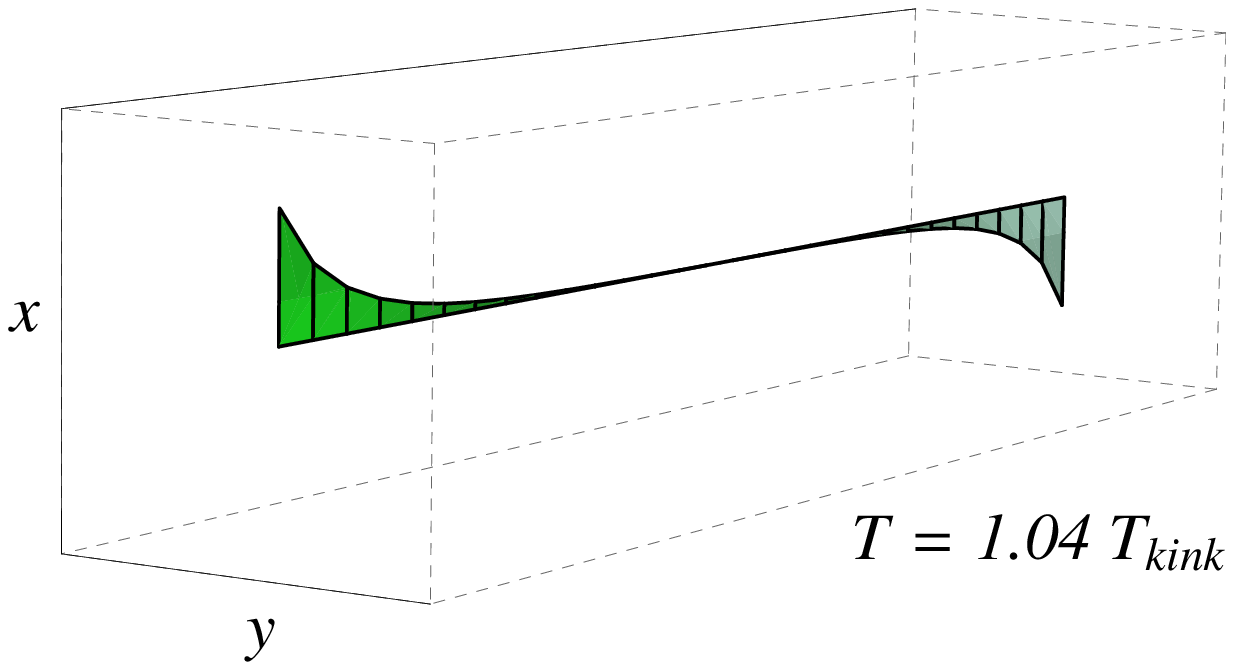}
\includegraphics[angle=0,width=\columnwidth]{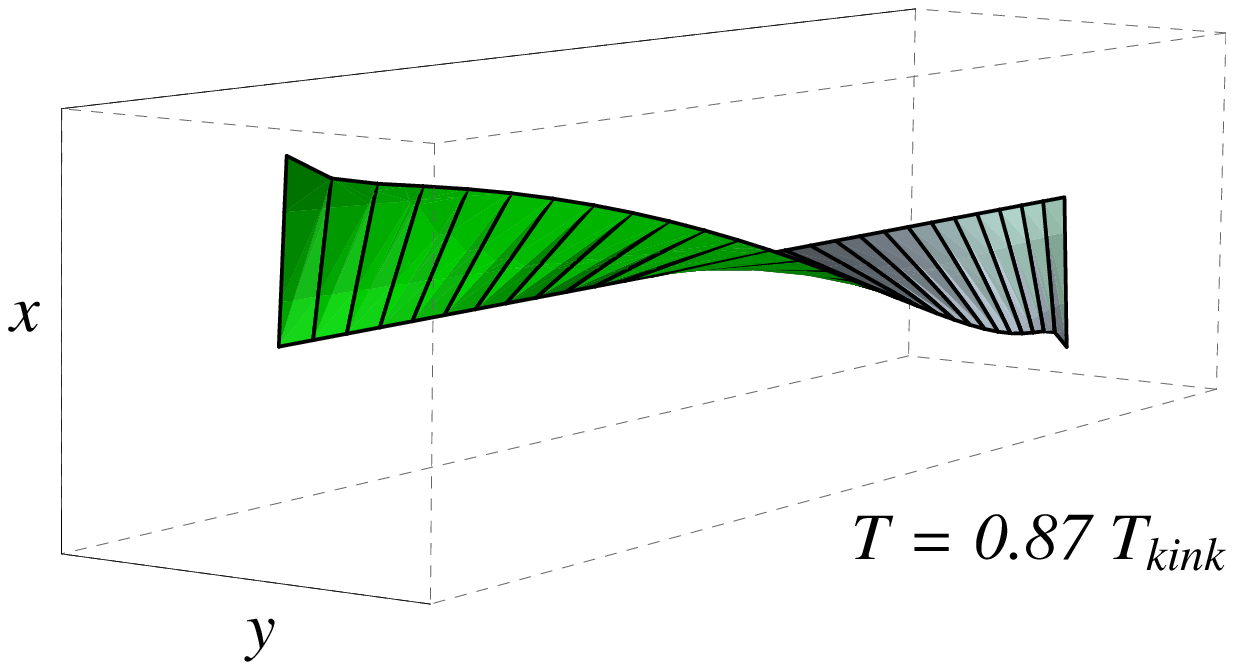}
\caption{(Color online) Renderings of the moments ($d=3$, $N=30$, $\alpha=\pi$) for temperatures above (left) and below (right) the temperature where a kink occurs in the Casimir force.}
\label{lattice_twist_illustration}
\end{figure*}

If we define ${\bf m}_0={\bf m}_{N+1}=0$ for notational convenience, then the function
\begin{equation}
f(\{{\bf m}_i\},N)=\sum_{i=1}^N\left[\frac{1}{2}{\bf m}_i\cdot{\bf H}_i-\frac{1}{\beta}\ln\left(I_0\left(\beta mH_i\right)\right)\right] \label{latticefreeenergy}
\end{equation}
may be regarded as the total free energy functional of the system, because
minimizing with respect to $\{{\bf m}_i\}$ yields the self-consistency
conditions (\ref{latticeeom}) (See Appendix \ref{latticefreeenergyappendix}). In
order to compute the Casimir force on the system, we must also find the free
energy per site in the bulk, i.e. when the system is very large. In that case,
the moments $\{{\bf m}_i\}$ will all be identical (at least near the center) and
Eq. (\ref{latticeeom}) tells us that the equation
$m_{i,\textrm{bulk}}=mR(2dm\beta Jm_{i,\textrm{bulk}})$ determines their common
magnitude $m_{i,\textrm{bulk}}$. For $T>dJm^2/k_B$, the only solution is
$m_{i,\textrm{bulk}}=0$ while for $T<dJm^2/k_B$, there is a non-zero solution.
Thus, the bulk system exhibits a transition at $T_{c,\textrm{bulk}}=dJm^2/k_B$
between an ordered ($T<T_{c,\textrm{bulk}}$) and a disordered phase.

Letting $a$ be the lattice spacing along the finite dimension of the system, the bulk free energy density is
\begin{equation}
f_b=\frac{1}{a}\left[dJm_{i,\textrm{bulk}}^2-\frac{1}{\beta}\ln\left(I_0\left(2dm\beta Jm_{i,\textrm{bulk}}\right)\right)\right]
\end{equation}
and the Casimir force will be computed as

\begin{figure}[hb]
\includegraphics[angle=0,width=\columnwidth]{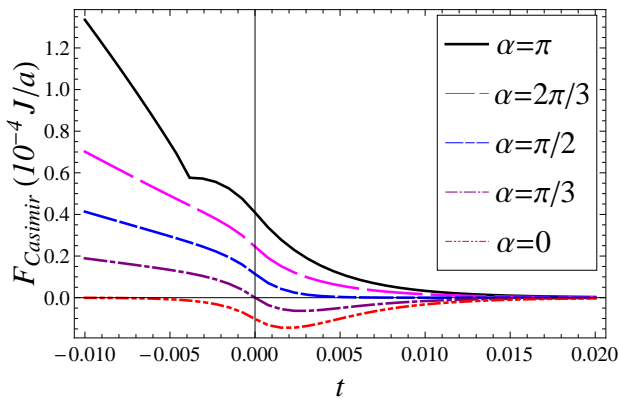}
\caption{(Color online) The Casimir force in the lattice model ($d=3$, $N=30$) as a function of reduced temperature $t=(T-T_{c,\textrm{bulk}})/T_{c,\textrm{bulk}}$ for various values of the twist angle, $\alpha$.}
\label{lattice_casimir_force}
\end{figure}

\begin{multline}
F_{\textrm{Casimir}}=-\frac{\partial}{\partial L}\left(f-Lf_b\right)\\
\approx f_b-\frac{f(N+1)-f(N)}{a}.
\end{multline}

We now restrict our attention to the case $d=3$ and investigate the system numerically. This amounts to solving the simultaneous Eqs. (\ref{latticeeom}) subject to particular boundary conditions on ${\bf m}_1$ and ${\bf m}_N$. We are interested in how the behavior of the system depends on the twist angle, $\alpha$.

When the twist angle is zero, the Casimir force is purely attractive (i.e, negative), as expected for matching boundary conditions. These features are illustrated in the plot of Casimir force versus reduced temperature (Fig. 2). As the twist is increased, a low temperature region of repulsive Casimir force emerges. At a twist angle of $\pi/2$, the Casimir force becomes purely repulsive. When the system nears anti-symmetric boundary conditions, $\alpha=\pi$, the Casimir force develops a kink at a temperature $T_{\textrm{kink}}<T_{c,\textrm{bulk}}$.

The nature of the kink will be discussed at greater length in a subsequent section, but we begin to understand it from the renderings (Fig. \ref{lattice_twist_illustration}). Below the kink temperature, the moments achieve the twist of $\pi$ by rotating about the axis while maintaining almost their full length $m$. Above the kink temperature, the moments all reside in a plane, and the twist is localized to the center of the system, where the magnetization has shrunk to zero. The nearest-neighbor interaction between moments imposes a free energy penalty for both rotating with respect to neighbors and varying in length. The transition indicates the point at which these penalties trade off in dominance.

\section{The Casimir force in the Ginzburg-Landau mean-field theory of the three-dimensional XY model}
\label{GLMFmodel}

We now consider the continuous analogue of this system in the Ginzburg-Landau mean-field theory. The order parameter of the system is the magnetization profile ${\bf m}(z)$, where $z$ is the finite dimension of the system. The behavior of the system is found by minimizing the free energy functional (per unit area),
\begin{multline}
{\cal F}\left[ {\bf m};t,L\right]=\int_{-L/2}^{L/2} dz\,\left[\frac{b}{2}\left|\frac{d{\bf m}}{dz}\right|^2+\frac{1}{2}at\left|\textbf{m}\right|^2\right.\\
\left.+\frac{1}{4}g\left|{\bf m}\right|^4\right],
\label{LGenergyfunctional}
\end{multline}
with respect to ${\bf m}$ and subject to certain boundary conditions. The quantity $t$ represents the reduced temperature.

Switching to polar coordinates,
\begin{equation}
{\bf m}(z)=\left(\Phi(z)\cos\varphi(z),\Phi(z)\sin\varphi(z)\right),
\end{equation}
the free energy functional is rewritten as
\begin{multline}
{\cal F}\left[\Phi,\varphi;t,L\right]=\int_{-L/2}^{L/2} dz\,\left[\frac{b}{2}\left(\frac{d\Phi}{dz}\right)^2+\frac{b}{2}\Phi^2\left(\frac{d\varphi}{dz}\right)^2\right.\\
\left.+\frac{1}{2}at\Phi^2+\frac{1}{4}g\Phi^4\right]. \label{LGfreeenergy}
\end{multline}

Minimization with respect to $\varphi(z)$ gives
\begin{equation}\label{varphiEq}
\frac{d}{dz}\left[\Phi^2{\varphi'}\right]=0
\end{equation}
which leads to
\begin{equation}
\Phi(z)^2 \left(\frac{d\varphi}{dz}\right)=P_\varphi
\label{Pvarphidef}
\end{equation}
with an integration constant $P_\varphi$, independent of $z$, which roughly indicates the degree of twisting in the system. The condition from minimizing with respect to $\Phi$ is similarly computed as
\begin{equation}
b\frac{d^2\Phi}{dz^2}=b\Phi\left(\frac{d\varphi}{dz}\right)^2+at\Phi+g\Phi^3
\label{PhiEq}
\end{equation}
or, with the identification (\ref{Pvarphidef}),
\begin{equation}
b\frac{d^2\Phi}{dz^2}=b\frac{P^2_\varphi}{\Phi^3}+at\Phi+g\Phi^3.
\label{Phieom}
\end{equation}
The problem is now that of solving Eq. (\ref{Phieom}) subject to twisted boundary conditions:
\begin{eqnarray}
&\varphi(\pm L/2)=\pm \alpha/2,\nonumber\\
&\Phi(\pm L/2) = \infty,
\label{boundaryconditions}
\end{eqnarray}
i.e. where the moments at the boundaries are twisted by an angle $\alpha$ relative to one another.

Note that, because of reflection symmetry in Eq. (\ref{Phieom}) and the boundary conditions imposed on $\Phi$, we have that $\Phi(z)=\Phi(-z)$ and, thus, $\Phi'(z)=-\Phi'(-z)$, whence $\Phi'(0)=0$. From the symmetry of Eq. (\ref{Pvarphidef}) one concludes $\varphi(z)=-\varphi(-z)$ which leads to $\varphi(0)=0$.

Multiplying (\ref{Phieom}) by $d\Phi/dz$ and integrating with respect to $z$, we find a first integral
\begin{equation}\label{solPhiprime}
{P}_\Phi=-\frac{1}{2}b \left[\frac{P_\varphi^2}{\Phi^2}+\left(\frac{d\Phi}{dz}\right)^2 \right]  +  \frac{1}{2} a\, t\, \Phi^2+\frac{1}{4} g\, \Phi^4,
\end{equation}
with $P_\Phi$ being another integration constant independent of $z$. Let $\Phi_0\equiv\Phi(z=0)$ be the amplitude of the order parameter at the center of the interval. Then, taking into account that $\Phi'(0)=0$ one can conveniently express $P_\Phi$ as
\begin{equation}\label{PhiPhinode}
{P}_\Phi=-\frac{1}{2}b\frac{P_\varphi^2}{\Phi_0^2}  +  \frac{1}{2} a\, t\, \Phi_0^2+\frac{1}{4} g\, \Phi_0^4,
\end{equation}
from which it follows that
\begin{multline}\label{Phifirstintegral}
\left(\frac{d\Phi}{dz}\right)^2=P^2_\varphi\left(\frac{1}{\Phi_0^2}-\frac{1}{\Phi^2}\right)+\hat{a}t\left(\Phi^2-\Phi_0^2\right)\\
+\frac{\hat{g}}{2}\left(\Phi^4-\Phi_0^4\right),
\end{multline}
where
\begin{equation}\label{hatvar}
\hat{a}=\frac{a}{b}, \qquad \hat{g}=\frac{g}{b}.
\end{equation}
The last result allows us to express the boundary conditions as
\begin{widetext}
\begin{eqnarray}
\frac{L}{2}=\int_0^{L/2} dz=\int_{\Phi_0}^{\infty}d\Phi\,\frac{dz}{d\Phi}
=\int_{\Phi_0}^{\infty}d\Phi\,\frac{1}{\sqrt{P^2_\varphi\left(\Phi_0^{-2}-\Phi^{-2}\right)+\hat{a}t\left(\Phi^2-\Phi_0^2\right)+\frac{\hat{g}}{2}\left(\Phi^4-\Phi_0^4\right)}} \label{lengthcondition} \\
{\rm and}\qquad\frac{\alpha}{2}=\int_0^{L/2} dz\,\frac{d\varphi}{dz}=P_\varphi\int_{\Phi_0}^\infty \frac{d\Phi}{\Phi^2}\frac{1}{\sqrt{P^2_\varphi\left(\Phi_0^{-2}-\Phi^{-2}\right)+\hat{a}t\left(\Phi^2-\Phi_0^2\right)+\frac{\hat{g}}{2}\left(\Phi^4-\Phi_0^4\right)}}. \label{twistcondition}
\end{eqnarray}
\end{widetext}
These equations relate the integration constants, $P_\varphi$ and $\Phi_0$, to the system's external parameters, $L$ and $\alpha$.

The stress tensor operator for a system with the free energy functional
(\ref{LGenergyfunctional}) is \cite{EiS94}
\begin{multline}\label{STO}
T_{k,l} = b \frac{\partial \mathbf{m}}{\partial x_k}\frac{\partial \mathbf{m}}{\partial x_l}\\
-\delta_{k,l} \left\{ \frac{1}{2} b \left[ {\Phi'}^2+\Phi^2{\varphi'}^2\right]+\frac{1}{2}at\Phi^2+\frac{1}{4}g\Phi^4\right\}\\
-b\left[\frac{d-2}{4(d-1)}+O\left(g^3\right) \right] \left[\frac{\partial^2}{\partial x_k \partial x_l} -\delta_{k,l}\nabla^2\right] \Phi^2.
\end{multline}
Calculating the $\langle T_{z,z} \rangle$ component, one obtains
\begin{equation}\label{STOfilm}
\langle T_{z,z} \rangle =\frac{1}{2}b\left[\left(\frac{d\Phi}{dz}\right)^2 +
\frac{P^2_\varphi}{\Phi^2}\right] - \frac{1}{2}at\Phi^2 - \frac{1}{4}g\Phi^4
\end{equation}
which, according to the general theory, is a $z$-independent quantity
equal to the pressure $-\partial f^{(\tau)}(T,L)/\partial L$ between  the
plates confining a fluctuating medium
\cite{K97,EiS94}. In our case, we see that
\begin{equation}
\langle T_{z,z} \rangle\equiv -P_\Phi,
\end{equation}
by (\ref{solPhiprime}). From Eqs. (\ref{PhiPhinode}) and (\ref{STOfilm}) one
observes that the Casimir force (the excess pressure over the bulk one) in this
system is
\begin{multline}\label{Casimir}
F_{\rm Casimir}(t,L)=-\left[-\frac{1}{2}b\frac{P_\varphi^2}{\Phi_0^2}  +  \frac{1}{2} a\, t\, \Phi_0^2\right.\\
\left.+\frac{1}{4} g\, \Phi_0^4 +\frac{1}{4g}(at)^2 \theta(-t)\right]
\end{multline}
where $\theta(x)$ is the Heaviside step function. Here, we have taken into account that the bulk free energy density $f_b$ for the system is $f_b(t<0)=-(a t)^2/4g$ while $f_b(t>0)=0$. As expected, the Casimir force is sensitive to the boundary conditions, through the quantities $P_\varphi$ and $\Phi_0$.

It is easy to show that $F_{\rm Casimir}(t,L)$  obeys the expected scaling. Indeed, in terms of the variables
\begin{multline}
z=L \zeta,\qquad \Phi=\sqrt{\frac{2}{\hat{g}}}X_\Phi L^{-1},\qquad \Phi_0=\sqrt{\frac{2}{\hat{g}}}X_0 L^{-1},\\
P_\varphi=\frac{2}{\hat{g}} X_\varphi L^{-3},\qquad \hat{a}t= x_t L^{-2}
\label{scalingvariables}
\end{multline}
the Casimir force reads (note that $\zeta$, $x_t$, etc. are all dimensionless)
\begin{equation}\label{cas}
F_{\rm Casimir}(t,L)=\frac{b}{\hat{g}}L^{-4}X_{\rm Cas}^{(\alpha)}(x_t),
\end{equation}
where
\begin{equation}\label{casscalingfunction}
X_{\rm Cas}^{(\alpha)}(x_t)=\left\{ \begin{array}{cc}
                             X_\varphi^2 / X_0^2- X_0^2\left(x_t+X_0^2 \right),  & x_t \ge 0 \\
                              X_\varphi^2 / X_0^2- \left(\frac{1}{2} x_t+X_0^2\right)^2, & x_t \le 0
                            \end{array} \right. .
\end{equation}
Taking into account that mean-field theories for short-ranges systems are effective $d=4$ theories, one concludes that Eq. (\ref{cas}) is in full agreement with the expected scaling behavior (\ref{CasGeneral}) of the Casimir force. From Eqs. (\ref{cas}) and (\ref{fssclowTas}) one can derive the low temperature asymptotic behavior of the scaling function $X_{\rm Cas}^{(\alpha)}(x_t)$ of the Casimir force. We find
\begin{equation}\label{casscalingfunctionasymp}
X_{\rm Cas}^{(\alpha)}(x_t)\simeq \frac{1}{2}  \alpha^2 |x_t|X_{\rm C},\; x_t\to-\infty,
\end{equation}
with $X_{\rm C}$ a constant. Appendix \ref{asymptotic} contains a derivation of the expression in  (\ref{casscalingfunctionasymp}). In that Appendix, we obtain the asymptotic expression for $X_{\rm Cas}^{(\alpha)}(x_t)$
\begin{equation}\label{casscalingfunctionasympXY}
X_{\rm Cas}^{(\alpha)}(x_t)\simeq \frac{1}{2}  \alpha^2 \left[|x_t|+4
 \sqrt{2|x_t|}+\frac{1}{2} \left(48-3 \alpha ^2\right)\right],
\end{equation}
when $x_t\to-\infty$. Note that Eq. (\ref{casscalingfunctionasympXY}) implies
that $X_{\rm C}=1$ for the $XY$ mean-field model.

One can further simplify (\ref{casscalingfunction}) by introducing the convenient combinations of scaling variables
\begin{equation}\label{newvarscaling}
\tau=x_t/X_0^2, \qquad \mbox{and} \qquad p=X_\varphi/X_0^3.
\end{equation}
Then the scaling function of the Casimir force reads
\begin{equation}\label{casscalingfunctionpandtau}
X_{\rm Cas}^{(\alpha)}(\tau)=\left\{ \begin{array}{cc}
                             X_0^4[p^2- \left(1+\tau \right)],  & \tau \ge 0 \\
                              X_0^4[p^2- \left(1+\tau/2 \right)^2], & \tau \le 0
                            \end{array} \right. .
\end{equation}
Eqs. (\ref{lengthcondition}) and (\ref{twistcondition}) then become
\begin{equation}\label{xoonpt}
X_0=\int_1^\infty \frac{dx}{\sqrt{(x-1)[x^2+x(1+\tau)+p^2]}},
\end{equation}
and
\begin{equation}\label{alphafinal}
\alpha= 2p \;X_0^3\int_0^{1/2} \frac{d\zeta}{X_\Phi^2(\zeta)}.
\end{equation}
\begin{figure}[ht]
\includegraphics[angle=0,width=\columnwidth]{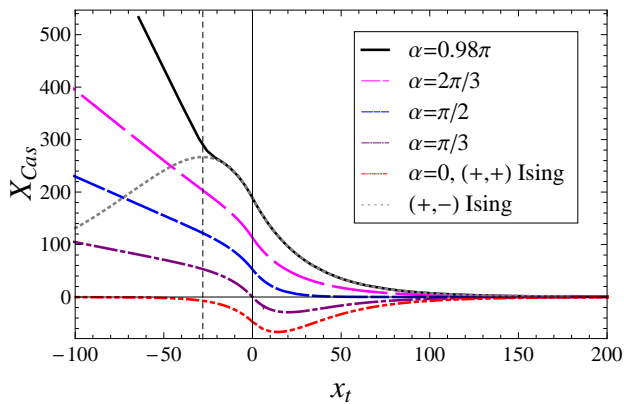}
\caption{(Color online) A plot of the dimensionless scaling function for the Casimir force,
$X_\textrm{Cas}$, versus $x_t$ (also dimensionless) for several values of
$\alpha$. The dotted curve is the Casimir scaling function in the Ising-like
case of a critical fluid under $(+,-)$ boundary conditions. For $x_t$ above a
certain value, this curve coincides with the one for the model studied here when
twisted by an angle $\alpha\approx\pi$.}
\label{plot:Casimir}
\end{figure}
\begin{figure}[ht]
\includegraphics[angle=0,width=\columnwidth]{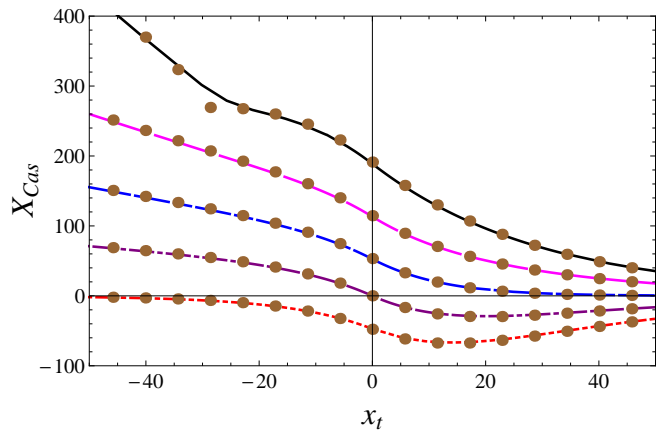}
\caption{(Color online) The scaling functions for the Casimir force in the Ginzburg-Landau
model (solid curves), overlaid with those from the lattice model with $N=50$
(data points), for several values of $\alpha$ (from top to bottom:
$\alpha=0.98\pi$, $\alpha=2\pi/3$, $\alpha=\pi/2$, $\alpha=\pi/3$, $\alpha=0$).}
\label{latticevslg_casimir_force}
\end{figure}

\begin{figure}[hb]
\includegraphics[angle=0,width=\columnwidth]{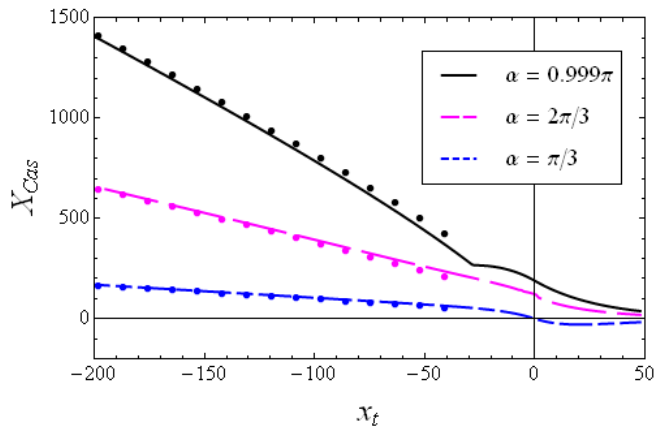}
\caption{(Color online) Casimir force curves in the Ginzburg-Landau model for several values of $\alpha$, overlayed with their respective asymptotic expressions (dotted) given by Eq. (\ref{casscalingfunctionasympXY}) as proven in Appendix C.}
\label{asymptotes}
\end{figure}

\begin{figure*}[ht]
\begin{center}
\includegraphics[angle=0,width=\columnwidth]{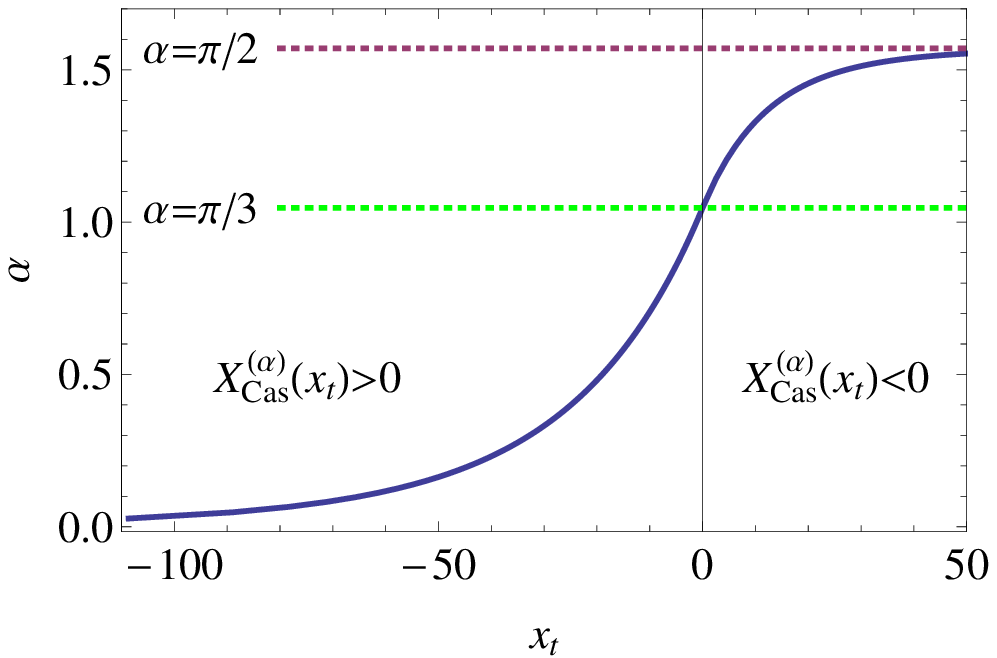}
\includegraphics[angle=0,width=\columnwidth]{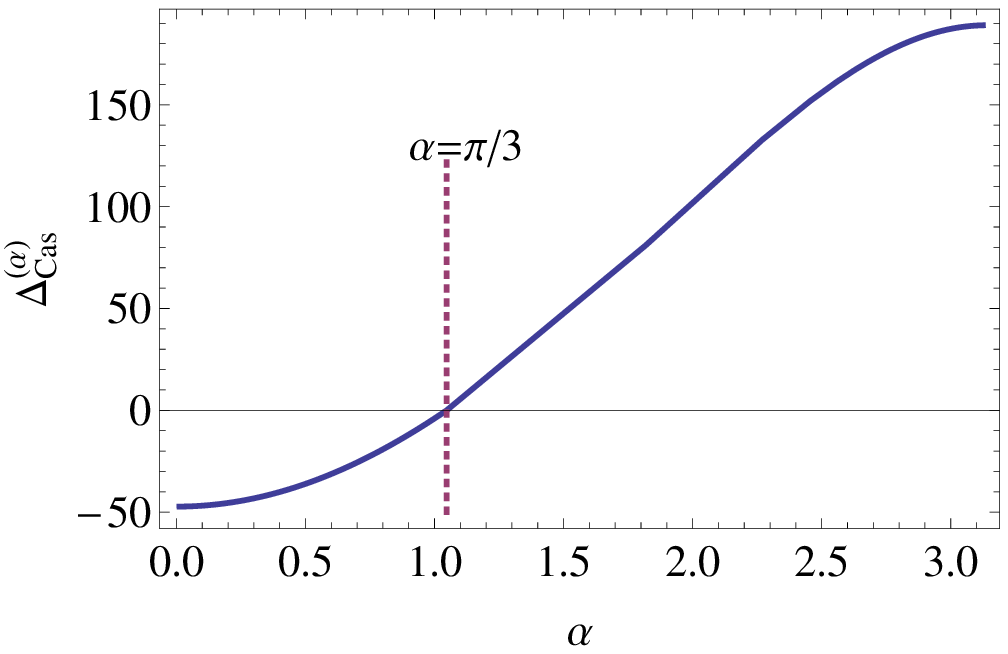}
\end{center}
\caption{(Color online) Left: The positions $x_{t,0}^{(\alpha)}$ of the zeros of the Casimir
force in the $(x_t,\alpha)$-plane. Right: The Casimir amplitude $\Delta_{\rm
Cas}^{(\alpha)}$ as a function of the twist angle $\alpha$. This curve has been
initially reported in \cite{K97} based on representation
(\ref{KrechAmplitudes}) derived there. The Casimir amplitude
changes sign at $\alpha=\pi/3$.}
\label{plot:Casimir_zeros_and_amplitudes}
\end{figure*}

In order to determine the Casimir force scaling function $X_{\rm Cas}^{(\alpha)}(x_t)$, all one needs to know is the behavior of $X_0=X_0(x_t|\alpha)$ and $X_\varphi(x_t|\alpha)=p X_0^3$ as functions of $x_t=\tau X_0^2$ at a given fixed value of the angle $\alpha$. For that, one has to solve Eqs. (\ref{xoonpt}) and (\ref{alphafinal}) after determining the function $X_\Phi(\zeta)$ from
\begin{multline}\label{zetaonxo}
\zeta=\frac{1}{2}\\
+\frac{1}{2 X_0}\int_{[X_\Phi(\zeta)/X_0]^2}^\infty\frac{dx}{\sqrt{(x-1)[x^2+x(1+\tau)+p^2]}}
\end{multline}
with $0<\zeta\le 1/2$, which directly follows from (\ref{Phifirstintegral}).
The detailed knowledge of the behavior of the phase angle profile
$\varphi(\zeta)$ is not needed. The analytical treatment of Eqs.
(\ref{xoonpt})-(\ref{zetaonxo}) is performed in Appendix \ref{GLMF}. The
numerical evaluation of the expressions derived there leads to the results for
the Casimir force presented in Fig. \ref{plot:Casimir}. The comparison shows
excellent agreement between the continuum and the lattice model results from the
previous section - see Fig. \ref{latticevslg_casimir_force}. In order to
demonstrate it, we scale the lattice results $(t, F_{\rm Casimir})$ to $(a_t N^2
t, a_F N^4 F_{\rm Casimir})$, where the scaling factors $a_t$ and $a_F$ are
determined by forcing the Casimir force with $\alpha=0$ to agree between the two
models. This was done numerically for $N=50$, where we find $a_t\approx 2.977$
and $a_F\approx 7.480\times 10^{-5}$. Fig. \ref{asymptotes} shows a comparison between the low temperature behavior of the Casimir force with the analytically derived asymptotic behavior reported in Eq. (\ref{casscalingfunctionasympXY}). We find that, for all $\alpha$, the asymptotic behavior is achieved for $x_t\lesssim -150$.

From Eq. (\ref{casscalingfunctionpandtau}) one can also infer some general
properties of the Casimir force. Taking into account that, at any fixed $x_t$
and $\alpha$, $p$ is a definite function of $x_t$ and $\alpha$, i.e. that
$p=p(\tau|\alpha)$ one can, e.g., determine the coordinates $x_{t,0}^\alpha$ of
the zeros for the Casimir force for a given angle $\alpha$. According to Eq.
(\ref{casscalingfunctionpandtau}) one has that $X_{\rm Cas}^{(\alpha)}=0$ for
$p(\tau|\alpha)=\sqrt{1+\tau}$, with $\tau\ge 0$ and for $p(\tau|\alpha)=
1+\tau/2$ when $-2\le\tau\le 0$. A plot of the positions of these zeros in the
$(x_t,\alpha)$-plane  is presented in Fig.
\ref{plot:Casimir_zeros_and_amplitudes}.

The figure demonstrates how, by changing, e.g., the twist angle $\alpha$, one
can, at a given temperature $t$, make the Casimir force either repulsive or
attractive. For $0<\alpha<\pi/2$, this can also be achieved by changing the
temperature, i.e. the scaling variable $x_t$, at a given fixed value of
$\alpha$. We also conclude that, when $\alpha \to 0$, the position of the zero
value of the Casimir force approaches $-\infty$. This implies that when
$\alpha=0$, the Casimir force will be attractive for all temperatures. Actually,
for $\alpha=0$, $X_{\rm Cas}^{(\alpha)}(x_t)$ coincides with the known result
for the Ising model system \cite{K97}. The behavior of $X_{\rm Cas}^{(+,+)}(x)$
is shown as a thick black line in Fig. \ref{plot:Casimir}. These results are
briefly re-derived in Appendix \ref{GLMF} for the convenience of the reader.

When $\alpha\to\pi/2$, we observe in Fig. \ref{plot:Casimir_zeros_and_amplitudes} that $x_{t,0}^{(\alpha)}\to \infty$. Thus $\alpha>\pi/2$ implies that the Casimir force will be repulsive for all values of $x_t$. As $\alpha$ increases, the repulsive force becomes stronger. We see from Fig. \ref{plot:Casimir} that, when $\alpha=\pi/2$, the force is practically zero for all temperatures above the critical temperature of the finite system, while, for $\alpha>\pi/2$, it is repulsive in the whole temperature region. The cases $\alpha=2\pi/3$ and $\alpha=0.98 \pi$ illustrate these features in the figure. One observes numerically that, for $x_t>-10$, the $\alpha=0.98 \pi$ curve agrees with that of the mean-field Ising model with $(+,-)$ boundary conditions. At lower temperatures, there is an abrupt departure from the Ising model which will be discussed in the section to follow. The analytical expressions for the Ising model are known from \cite{K97}. For completeness, these results are recalled in Eq. (\ref{CasimiralphaeqpiKrech}) of Appendix \ref{GLMF}.

The behavior of the critical Casimir force, $\Delta_{\rm Cas}^{(\alpha)}=X_{\rm
Cas}^{(\alpha)}(x_t=0)/3$, as a function of $\alpha$, is illustrated in Fig.
\ref{plot:Casimir_zeros_and_amplitudes}. Note that the Casimir amplitude becomes
zero at $\alpha=\pi/3$, so that the Casimir force for $\alpha=\pi/3$ changes its
sign at $x_t=0$. This was initially reported in \cite{K97} and may also be seen
in Fig. \ref{plot:Casimir_zeros_and_amplitudes}. Note that in \cite{K97}
different type of parametrization of the amplitude and phase profiles is used---they are parametrized via the Casimir amplitudes.  This led to a restriction
of the results presented to the critical temperature only. For example, for the
determination of the Casimir amplitudes in \cite{K97} one has to solve, in our
notations, the following system of equations (see Eq. (3.16) in \cite{K97})
\begin{subequations}
\label{KrechAmplitudes}
\begin{equation}
 X_0=\int_1^\infty[x^3-1+(x-1)X_0^{-4}\Delta_{\rm Cas}^{(\alpha)}]^{-1/2}dx
\end{equation}
and
\begin{eqnarray}
\alpha &=& \sqrt{1+X_0^{-4} \Delta_{\rm
Cas}^{(\alpha)}} \\
&&\int_1^\infty x^{-1}[x^3-1+(x-1)X_0^{-4}\Delta_{\rm
Cas}^{(\alpha)}]^{-1/2}dx. \nonumber
\end{eqnarray}
\end{subequations}

\section{The transition at $\alpha=\pi$}
\label{alphaequalpicase}

The case $\alpha=\pi$ warrants special investigation because it features behavior reminiscent of a  phase transition. As mentioned in Section 3, the high temperature behavior of the system at $\alpha\approx\pi$ tracks that of the Ising model. However, we find that a kink develops in all quantities in the system at a temperature $t_{\textrm{kink}}$ below the bulk critical temperature of the system, and the system changes its character at this temperature. The lattice model also featured such a kink. In Section 2, we illustrated how the lattice system switches from a ``rotational'' state below the kink temperature to a ``planar'' state above it. We note that this phase transition-like behavior exists only in the finite system under the given boundary conditions and not in a thermodynamic sense. Below the transition temperature, the two coexisting phases are the rotational states with rotation plus or minus $\pi$. There is spontaneous symmetry breaking when the system orders in one of them. Above the critical temperature, there is a single state -- the ``planar'' one.

\begin{figure}[ht]
\includegraphics[angle=0,width=\columnwidth]{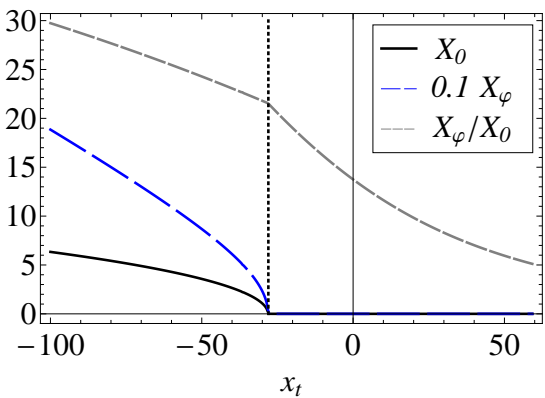}
\caption{(Color online) The dimensionless quantities $X_0$ (solid), $X_\varphi/10$ (long dashing; scale reduced for ease of plotting) and $X_\varphi/X_0$ (short dashing) as functions of $x_t$ when $\alpha\approx\pi$. The vertical dotted line indicates $x_t=x_{t,\rm kink}$.}
\label{quantitykinkplots}
\end{figure}

The system incurs free energy penalties when adjacent moments vary in length or direction. We see how each type of state mentioned would extremize the energy. The moments in the planar state minimize rotation: they reside in a plane and shorten to a length of zero at the center of the interval, where an abrupt reversal of direction occurs. The moments in the rotational state minimize length variation while gradually rotating from one end of the interval to the other. The temperature at which the kink occurs is the point at which the two energy penalties trade off in dominance. According to this description, the planar state is characterized by $X_0=X_\varphi=0$. Indeed, this is what comes out of the Ginzburg-Landau model above the kink temperature (see Fig. \ref{quantitykinkplots}). While these quantities vanish at high temperature, their ratio $X_\varphi/X_0$ remains non-zero at all temperatures.

In order to determine the kink temperature, we solve Eqs. (\ref{lengthcondition}) and (\ref{twistcondition}) simultaneously in the vicinity of that transition point, i.e. with $X_0=0$ and $X_\varphi/X_0$ finite but unknown. We find numerically (see Appendix \ref{tkinkappendix})
\begin{equation}
x_{t,\textrm{kink}}\approx -28.1099. \label{kinktemp}
\end{equation}
Additionally, series expanding the conditions (\ref{lengthcondition}) and (\ref{twistcondition}), we can find a first approximation to $X_0(x_t)$ in the vicinity of the kink. The result is that
\begin{equation}
X_0(x_t)\propto(x_{t,\textrm{kink}}-x_t)^{1/2}
\end{equation}
which agrees with standard mean field results for, e.g. the magnetization of a ferromagnet.

We now present another, more transparent, analysis of this transition. We claim that, at high temperatures, the system's only energy extremum is the planar state, while, at low temperatures, the system has access to the planar state, as well as two energetically equivalent rotational states with a gradual turn of either $\pi$ or $-\pi$. As previously described, the system favors the rotational state at low temperatures and ignores the planar state. Such a situation is commonly found in Ginzburg-Landau models, where the free energy has terms quartic and quadratic in a variable or field of interest. Depending on the coefficient of the quadratic term, there will be one fourth-order minimum at the origin or else a maximum at the origin and minima elsewhere. The planar state can be either an energy minimum or an energy maximum, while the rotational states are the symmetry-breaking minima that appear at sufficiently low temperature. By employing an approximation, we will show that this description fits the present system.

It is useful to turn to the scaling variables (\ref{scalingvariables}), in which the free energy functional, Eq. (\ref{LGfreeenergy}), becomes
\begin{multline}
{\cal F}=\frac{b}{\hat{g}} \frac{1}{L^3}\int_{-1/2}^{1/2} d\zeta \,\left[\left(\frac{dX_\Phi}{d\zeta}\right)^2+\frac{X_\varphi^2}{X_\Phi^2}\right.\\
\left.+x_t  X_\Phi^2+ X_\Phi^4\right]
\end{multline}
while the amplitude equation (\ref{Phieom}) reads
\begin{equation}
\frac{d^2X_\Phi}{d \zeta^2}=\frac{X^2_\varphi}{X_\Phi^3}+x_t X_\Phi+2X_\Phi^3.
\label{PhiEqScaled}
\end{equation}
Due to the boundary conditions (\ref{boundaryconditions}) the amplitude profile near the edges, $X_\Phi(\zeta\approx\pm 1/2)$, is largely temperature independent, i.e. it is almost the same above and below the transition temperature. The behavior near $\zeta=0$ must therefore account for the physics of the transition. Near the transition temperature, we have $X_0\approx 0$ and thus $X_\Phi(\zeta)\ll 1$ for $\zeta\approx 0$. Then Eq. (\ref{PhiEqScaled}) reduces to
\begin{equation}
\frac{d^2 X_\Phi}{d\zeta^2}=\frac{X^2_\varphi}{X_\Phi^3}
\end{equation}
with solution
\begin{equation}
X_\Phi(\zeta)=\sqrt{X_0^2+\left(\frac{X_\varphi}{X_0}\right)^2\zeta^2}\equiv\sqrt{X_0^2+M^2\zeta^2}, \label{asymptamp}
\end{equation}
defining $M\equiv X_\varphi/X_0$.

The free energy integral can now be computed in closed form using the asymptotic expression (\ref{asymptamp}). The expression for $X_\Phi(\zeta)$ is only valid up to some cutoff $\zeta=Y<1/2$, but the free energy from the edges of the interval ($|\zeta|>Y$) will not contribute to the behavior of the system, provided that $Y$ is large enough. The result of the integral is
\begin{multline}
{\cal F}\approx\frac{2bY}{\hat{g}L^3}\left[X_0^4 + \left(x_t+\frac{2}{3} Y^2 M^2\right)X_0^2\right.\\
\left. + \left(\frac{1}{5} Y^4 M^4 + \frac{1}{3}x_t Y^2 M^2  + M^2\right)\right] \label{approxfreeenergy}
\end{multline}
which is quartic in $X_0$. Recall that $X_0$ is the (scaled) amplitude of the order parameter at the center of the interval. In the planar state, $X_0=0$, while rotational states have $X_0>0$. Therefore we expect to always see an extremum at $X_0=0$, corresponding to the planar state, which will be a free energy minimum at high temperature and a maximum at low temperature. When the planar state is a maximum, two minima (rotational states) with $|X_0|>0$ should emerge. Indeed, this behavior is clear from the form of (\ref{approxfreeenergy}). The position of the non-zero minimum is found to be
\begin{equation}
X_0=\pm\sqrt{-\frac{1}{2}\left(x_t+\frac{2}{3} Y^2 M^2\right)}, \label{B0minimum}
\end{equation}
which is only real for sufficiently low temperature: $x_t\le -2 Y^2 M^2/3<0$. The transition occurs when equality holds. This does not fully determine the temperature of the transition because both $Y$ and $M$ are functions of $x_t$. The additional constraints are afforded by matching the hyperbolic expression (\ref{asymptamp}) for $X_\Phi(\zeta)$ with another expression correct near the edge of the interval.

When $\zeta\approx \pm 1/2$, $X_\Phi\to\infty$ and, according to (\ref{PhiEqScaled}) the amplitude profile is determined by
\begin{equation}
\frac{d^2 X_\Phi}{d\zeta^2}=x_t X_\Phi + 2 X_\Phi^3
\end{equation}
which, with $x_t<0$, is solved by
\begin{equation}
X_\Phi(\zeta)=\sqrt{|x_t|}\,\csc\left[\sqrt{|x_t|}\left(\frac{1}{2}-|\zeta|\right)\right].
\label{largezeta}
\end{equation}
Now imposing the continuity of $X_\Phi(\zeta)$ and $X'_\Phi(\zeta)$ from (\ref{asymptamp}) and (\ref{largezeta}) at $\zeta=Y$ allows us to solve for the parameters at the transition point. Proceeding numerically, we find:
\begin{eqnarray}
&x_{t,\textrm{kink}}\approx-22.4587,\qquad M_{\textrm{kink}} \approx 19.4498,\nonumber\\
&X_{0,\textrm{kink}}=0,\qquad Y_{\textrm{kink}}\approx0.2984.
\end{eqnarray}
Compared to the exact numerical results obtained in Appendix
\ref{tkinkappendix}, these values are consistent as a first approximation, as
are the behaviors of $X_0(x_t)$ and $M(x_t)$. In particular, the leading order
contribution to $X_0(x_t)$ goes as $(x_{t,\textrm{kink}}-x_t)^{1/2}$, in
agreement with the power law previously found. Finally, it is easy to
show that at the kink temperature the rate of the change of the phase in the
middle of the system diverges. Indeed, from Eq. (\ref{Pvarphidef}) and using the
definitions (\ref{scalingvariables}) one obtains
\begin{equation}
 \left[\frac{X_\Phi(\zeta)}{X_0}\right]^2\frac{d\varphi(\zeta)}{d
\zeta}=\frac{X_\varphi}{X_0^2}=\frac{M}{X_0}.
\label{relMkink}
\end{equation}
Thus, in the limit $\zeta\to 0$ one derives at
$x_t=x_{t,\textrm{kink}}$ that $(d \varphi(\zeta)/ d\zeta)|_{\zeta=0}=M/X_0\to
\infty$, since $X_{0,\textrm{kink}}=0$. Therefore, at the kink temperature all
the change in the phase of the moments happens at the middle of the system,
where there length becomes zero. The phase of the moments jumps there from $0$
to $\pi$.

\section{Discussion and concluding remarks}
\label{discussion}

We have studied the $O(2)$ model in a three-dimensional film geometry, and find
identical predictions from lattice and continuum mean-field theories. Consistent
with systems of similar type, the Casimir force with symmetric boundary
conditions ($\alpha=0$) is attractive, while the Casimir force with
anti-symmetric boundary conditions ($\alpha=\pi$) is repulsive. In particular,
the critical Casimir force, i.e. $F_{\rm Casimir}$ at the bulk transition
temperature, changes from repulsive to attractive. This is a standard result,
but we also find intermediate scenarios when $0<\alpha<\pi$ which feature
critical Casimir forces, and scaling functions for the Casimir force
(illustrated in Fig. \ref{lattice_casimir_force}), different from those of the
symmetric and anti-symmetric cases. The
Casimir force may therefore be continuously adjusted at constant temperature by
varying the twist $\alpha$, or at constant twist by varying the temperature.

Additionally we find that, when the boundary conditions are perfectly anti-symmetric ($\alpha=\pi$), the system undergoes a phase transition at a temperature below the bulk critical temperature. We are able to understand this transition as a symmetry-breaking effect: at high temperatures the moments of the system are confined to a plane, while at low temperatures they rotate about the $z$-axis by either $\pi$ or $-\pi$ to satisfy the boundary conditions. The high temperature behavior tracks that of an Ising model whose order parameter is always simply up or down, but our system departs from that behavior at the transition point, once moments find it energetically favorable to rotate.

\acknowledgments{D. Dantchev acknowledges the partial financial support of
Bulgarian NSF, grant No. DO 171/08. J. Rudnick acknowledges partial support from the NSF through grant DMR-1006128.}

\appendix
\section{Lattice free energy}
\label{latticefreeenergyappendix}
We claim that
\begin{equation}
f(\{{\bf m}_i\},N)=\sum_{i=1}^N\left[\frac{1}{2}{\bf m}_i\cdot{\bf H}_i-\frac{1}{\beta}\ln\left(I_0\left(\beta mH_i\right)\right)\right]
\label{latticefreeenergyA}
\end{equation}
is the free energy functional of the lattice model considered in Section 2.
Indeed, here we will demonstrate that minimizing it with respect to the ${\bf
m}_i$ leads to the mean-field consistency equations
\begin{equation}
{\bf m}_i=m\frac{{\bf H}_i}{H_i}\frac{I_1(\beta mH_i)}{I_0(\beta mH_i)}.
\label{latticeeomA}
\end{equation}
Recall that ${\bf H}_i=J(2(d-1){\bf m}_i+{\bf m}_{i-1}+{\bf m}_{i+1})$ and that we take ${\bf m}_0={\bf m}_{N+1}=0$ for notational convenience. Differentiating,
\begin{multline}
0=\nabla_{{\bf m}_i} f=2J(d-1)\left[{\bf m}_i-m\frac{{\bf H}_i}{H_i}\frac{I_1(\beta mH_i)}{I_0(\beta mH_i)}\right]\\
+J\left[{\bf m}_{i-1}-m\frac{{\bf H}_{i-1}}{H_{i-1}}\frac{I_1(\beta mH_{i-1})}{I_0(\beta mH_{i-1})}\right]\\
+J\left[{\bf m}_{i+1}-m\frac{{\bf H}_{i+1}}{H_{i+1}}\frac{I_1(\beta mH_{i+1})}{I_0(\beta mH_{i+1})}\right] \label{energymincondition}
\end{multline}
for $i=2,\ldots,N-1$. If $i=1$ ($i=N$), we find the same condition with the second (third) term omitted. Defining
\begin{equation}
{\bf g}_i={\bf m}_i-m\frac{{\bf H}_i}{H_i}\frac{I_1(\beta mH_i)}{I_0(\beta mH_i)},\qquad {\bf g}_0={\bf g}_{N+1}=0,
\end{equation}
we may write Eq. (\ref{energymincondition}) as
\begin{equation}
0=\nabla_{{\bf m}_i} f=2J(d-1){\bf g}_i+J{\bf g}_{i-1}+J{\bf g}_{i+1}\label{freeenergymincondition}
\end{equation}
for each $i=1,\ldots,N$. In order to show that (\ref{latticeeom}) holds, we must show that (\ref{freeenergymincondition}) is only solved when ${\bf g}_i=0$ for all $i$. This is seen by writing the linear equations (\ref{freeenergymincondition}) in matrix form, i.e. $A{\bf g}=0$ with tridiagonal $N\times N$ matrix
\begin{equation}
A=J\left(\begin{array}{ccccc}
2(d-1) & 1      & 0      & \cdots & 0      \\
1      & 2(d-1) & 1      &        & \vdots \\
0      & 1      & \ddots &        &        \\
\vdots &        &        &        &        \\
       &        &        &        & 1      \\
0      & \cdots &        & 1      & 2(d-1)
\end{array}\right).
\end{equation}
whose determinant is computed as
\begin{multline}
\det A=\frac{J^N}{2\sqrt{(d-1)^2-1}}\\
\times\left[\left(d-1+\sqrt{(d-1)^2-1}\right)^{N+1}\right.\\
\left.-\left(d-1-\sqrt{(d-1)^2-1}\right)^{N+1}\right]
\end{multline}
which is non-zero for $d\ge 2$ and $N\ge 1$, so the only solution is ${\bf g}_i=0$, as desired.

\section{The amplitude and phase profiles and Casimir force derivation within the Ginzburg-Landau mean-field theory of the three-dimensional XY model}
\label{GLMF}

% Here the results from the file Elliptic Integrals 3.nb have been used.

In this appendix, we derive some analytical expressions needed for the numerical evaluation of Eq. (\ref{casscalingfunction}) for the scaling function of the Casimir force.
\begin{figure}[h!]
\includegraphics[angle=0,width=\columnwidth]{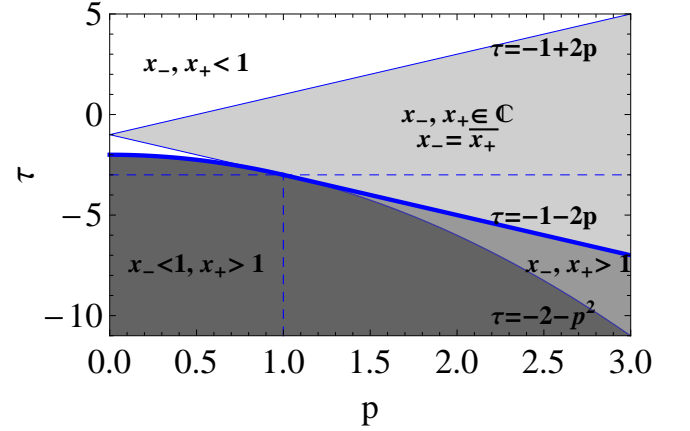}
\caption{(Color online) The loci of points in the $(p,\tau)$-plane for which the roots $x_\pm$ possess the properties discussed in the main text. When any of the roots approaches the thick blue line, $X_0\to\infty$. One observes that this is only possible  for $\tau<0$, i.e. when $\tau X_0^2=x_t\to -\infty$.}
\label{loci_roots}
\end{figure}

We start by determining the behavior of the amplitude profile $X_\Phi(\zeta)$ - see Eq. (\ref{zetaonxo}). In addition, we will also determine the phase angle profile $\varphi(\zeta)$. Note that, in terms of the scaling variables (\ref{scalingvariables}) and (\ref{newvarscaling}), we obtain the phase angle $\varphi(\zeta)$ as
\begin{equation}\label{varphionzeta}
\varphi(\zeta)= p \;X_0^3\int_0^\zeta \frac{d\zeta}{X_\Phi^2(\zeta)}.
\end{equation}
via Eq. (\ref{Pvarphidef}). Let
\begin{equation}\label{roots}
x_\pm=\frac{1}{2} \left[-(\tau +1)\pm\sqrt{(\tau +1)^2-4 p^2}\right]
\end{equation}
be the roots of the quadratic term in the square brackets in the denominator of (\ref{xoonpt}).

In order to perform the integration in (\ref{xoonpt}), where the integrand is a
positive function for all points from the integration interval, one needs to
know if these roots are real or complex (see Fig. 8). Thus, there are two subcases: {it A)}
the roots are real, and {\it B)} the roots are complex conjugates of each
other.

First consider the subcase

{\it A)} The roots $x_\pm$ are real.

In that case the positivity of the integrand implies $x_-<x_+<1$. Taking the
above into account and using the corresponding expression reported in \cite{GR},
\begin{multline}\label{GR_3.131.8}
\int_y^\infty \frac{dx}{\sqrt{(x-1)(x-x_+)(x-x_-)}}\\
=\frac{2}{\sqrt{1-x_-}}F\left(\arcsin \sqrt{\frac{1-x_-}{y-x_-}}, \sqrt{\frac{x_+-x_-}{1-x_-}} \right),
\end{multline}
provided $y \ge 1>x_+>x_-$. From Eq. (\ref{zetaonxo}), one obtains
\begin{multline}\label{dzeta}
\zeta-\frac{1}{2}=\frac{1}{X_0\sqrt{1-x_-}}
\\\times F\left[\arcsin \sqrt{\frac{1-x_-}{(X_\Phi/X_0)^2-x_-}}, \sqrt{\frac{x_+-x_-}{1-x_-}} \right],
\end{multline}
so that, for $\zeta=0$ with $X_\Phi(\zeta=0)=X_0$, it follows that
\begin{equation}\label{x0det}
X_0=\frac{2}{\sqrt{1-x_-}}K\left[\sqrt{\frac{x_+-x_-}{1-x_-}} \right].
\end{equation}
In (\ref{GR_3.131.8}) and (\ref{dzeta}), $F$ refers to the elliptic integral of the first kind, while $K$ in (\ref{x0det}) is the complete elliptic integral of the first kind.
Solving (\ref{dzeta}) for $X_\Phi$ one finds
\begin{multline}\label{Xphi}
X_\Phi^2\left(\zeta+\frac{1}{2}\right)\\
=X_0^2 \left[x_-+\frac{1-x_-}{{\rm sn}^2\left[\sqrt{1-x_-} \,X_0 \zeta, \sqrt{\frac{x_+-x_-}{1-x_-}}\right]}  \right],
\end{multline}
where ${\rm sn}$ denotes the corresponding sine-amplitude Jacobi elliptic function.
\begin{figure*}[ht]
\begin{center}
\includegraphics[width=\columnwidth]{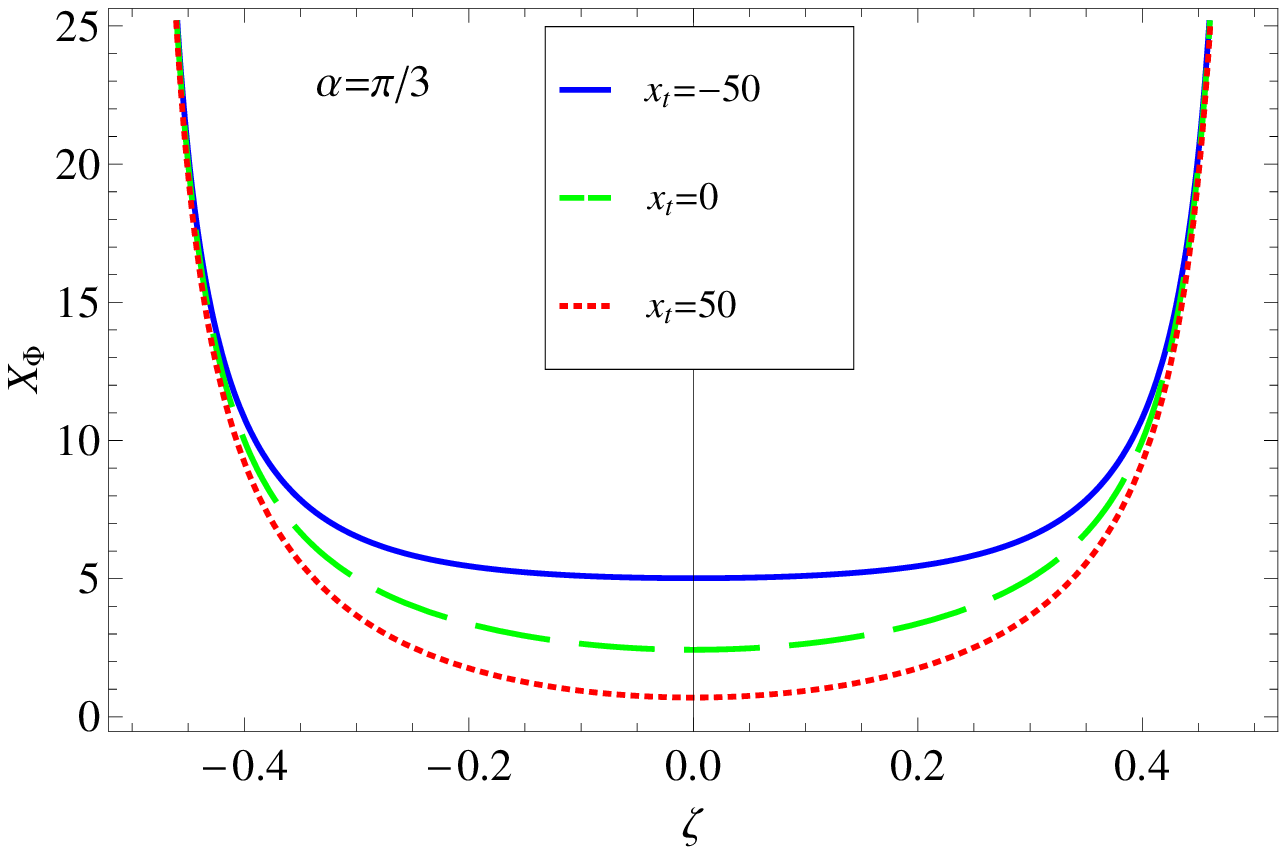}
\includegraphics[width=\columnwidth]{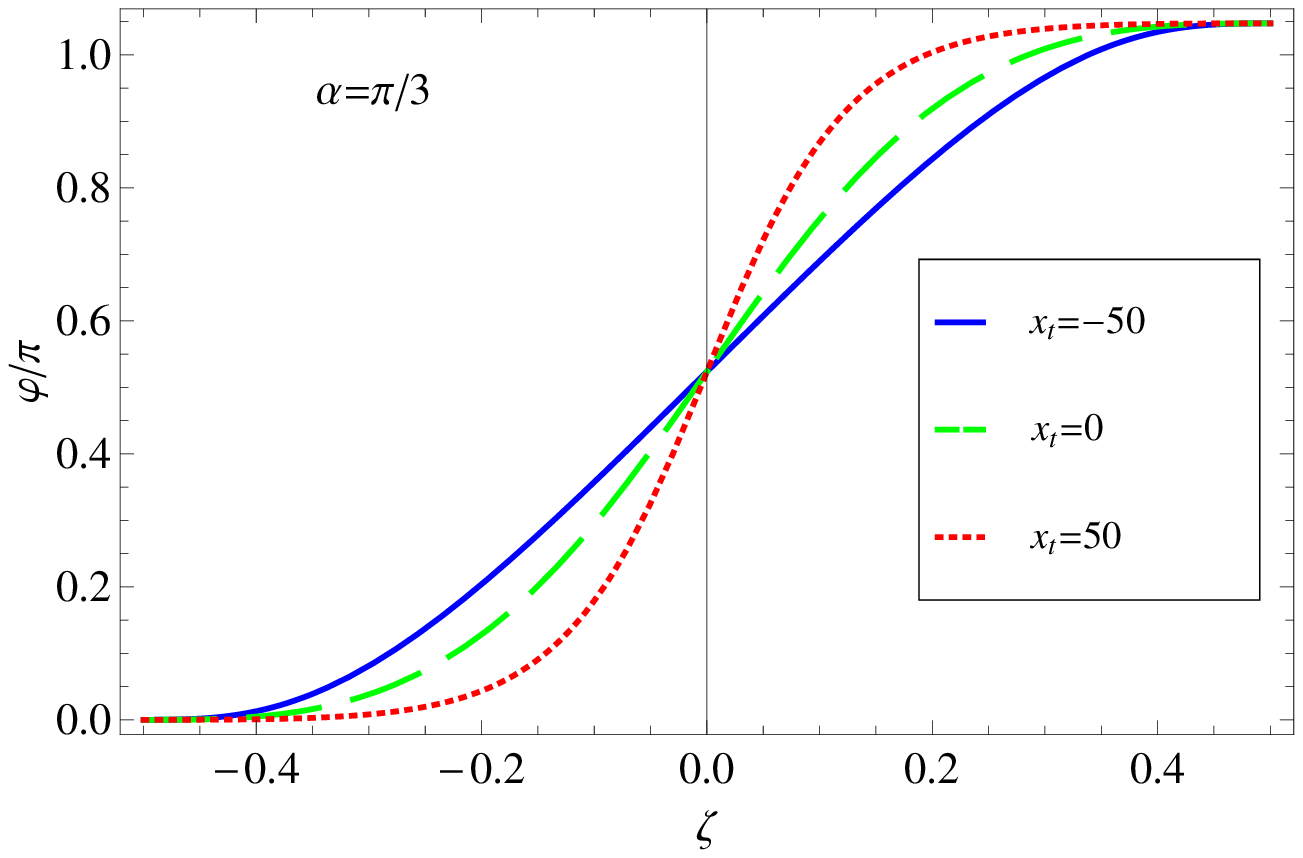}
\end{center}
\caption{(Color online) Plots of the amplitude profile and the angle of the order parameter for $\alpha=\pi/3$ and some choices of $x_t$. We observe that, when the temperature increases, the value of the amplitude in the middle of the system decreases. The twist of the local variables through the system spans over the total system almost uniformly for low temperatures, while for higher ones it concentrates more and more in the middle of the system where the amplitude is at its smallest values.}
\label{plot:profiles}
\end{figure*}
Finally, inserting (\ref{Xphi}) into (\ref{varphionzeta}) and (\ref{alphafinal}) and performing the integration, we arrive at
\begin{widetext}
\begin{eqnarray}
\varphi\left(\zeta+\frac{1}{2}\right)=\frac{\sqrt{|x_- x_+|} X_0}{x_-} \left\{\zeta-\frac{1}{X_0 \sqrt{1-x_-}}\Pi \left[\frac{x_-}{x_- -1}, {\rm am}\left(X_0 \sqrt{1-x_-}\,\zeta,\sqrt{\frac{x_+-x_-}{1-x_-}}\right),\sqrt{\frac{x_+-x_-}{1-x_-}} \right]\right\}, \label{phiangleonz} \\
{\rm and}\qquad\alpha=\frac{\sqrt{|x_- x_+|} X_0}{x_-}\left\{1-\frac{2}{X_0 \sqrt{1-x_-}} \Pi \left[\frac{x_-}{x_- -1},{\rm am}\left(\frac{1}{2} X_0 \sqrt{1-x_-},\sqrt{\frac{x_+-x_-}{1-x_-}}\right),\sqrt{\frac{x_+-x_-}{1-x_-}} \right]\right\}, \label{alphaasafunctionoftheroots}
\end{eqnarray}
\end{widetext}
where $\Pi(n,\phi,m)$ is the incomplete elliptic integral of the third kind and ${\rm am}(u,m)$ is the amplitude for Jacobi elliptic functions.

The relationship between $x_\pm$, $\tau$ and $p$ is straightforward. From (\ref{roots}) we have
\begin{equation}\label{relxtp}
\tau=-1-x_--x_+,\qquad p=\sqrt{|x_- x_+|}.
\end{equation}
Taking into account Eq. (\ref{x0det}) one can further simplify Eq. (\ref{alphaasafunctionoftheroots}) to
\begin{multline}\label{alphaasafunctionoftherootsfinal}
\alpha=\frac{\sqrt{|x_- x_+|} X_0}{x_-} \bigg\{1\\
-\frac{2}{X_0 \sqrt{1-x_-}} \Pi \left[\frac{x_-}{x_- -1},\sqrt{\frac{x_+-x_-}{1-x_-}} \right]\bigg\}.
\end{multline}
Here $X_0$, $x_{-}$ and $x_+$ are known functions of $\tau$ and $p$. Fixing, for instance, $\tau$, one can solve (\ref{alphaasafunctionoftherootsfinal}) numerically for $p$. Then, knowing $X_0$, the scaling function for the Casimir force is found from Eq. (\ref{casscalingfunctionpandtau}). These scaling functions are plotted in Fig. \ref{plot:Casimir}.

{\it B)} The roots $x_\pm$ are complex.

In this case, the roots are complex conjugates of each other, i.e.
$x_-=\overline{x_+}$. Taking this into account and using and using the
corresponding expression reported in \cite{BF}, we obtain
\begin{multline}\label{BF241.00}
\int_y^\infty \frac{dx}{\sqrt{(x-1)(x-x_+)(x-x_-)}}\\
=\frac{1}{\sqrt{r}} F\left[\arccos\left(\frac{y-1-r}{y-1+r}\right), w\right],
\end{multline}
where
\begin{eqnarray}\label{notations1}
r\equiv r(x_-,x_+)&=&\sqrt{(1-x_-)(1-x_+)} \nonumber\\
&=& \sqrt{2+\tau+p^2},
\end{eqnarray}
and
\begin{multline}\label{notations2}
w^2\equiv w^2(x_-,x_+) = \frac{1}{2}+\frac{\frac{x_- + x_+}{2}-1}{2
   \sqrt{(1-x_-) (1-x_+)}}\\
   = \frac{1}{2}\left(1-\frac{3+\tau}{2\sqrt{2+\tau+p^2}} \right).
\end{multline}
According to Eq. (\ref{zetaonxo}), the above implies that
\begin{multline}\label{cdzeta}
\zeta=\frac{1}{2}\\
+\frac{1}{2 X_0\sqrt{r}}F\left[\arccos\left(\frac{(X_\Phi/X_0)^2-1-r}{(X_\Phi/X_0)^2-1+r}\right), w\right],
\end{multline}
and, for $\zeta=0$ with $X_\Phi(\zeta=0)=X_0$, it follows that
\begin{equation}\label{cx0det}
X_0=\frac{2}{\sqrt{r}} K\left(w\right).
\end{equation}
Solving (\ref{cdzeta}) for $X_\Phi$ gives
\begin{multline}\label{cXphi}
X_\Phi^2\left(\zeta+\frac{1}{2}\right)\\
=X_0^2 \left[1-r+\frac{2 r}{1-{\rm cn}\left(2X_0\sqrt{r}\,\zeta,w\right)}\right],
\end{multline}
where ${\rm cn}$ denotes the corresponding cosine-amplitude Jacobi elliptic function.
Finally, inserting (\ref{cXphi}) in (\ref{varphionzeta}) and (\ref{alphafinal}) and performing the integration, one arrives at
\begin{multline}\label{cphiangleonz}
\varphi\left(\zeta+\frac{1}{2}\right) = \frac{p}{r^2-1} \sqrt{\frac{r}{1-w^2}}\\
\times\left\{\Pi\left[\left(\frac{r-1}{r+1}\right)^2,\frac{w}{\sqrt{w^2-1}}\right]\right.\\
\left.-\Pi\left[\left(\frac{r-1}{r+1}\right)^2, \frac{\pi}{2}-\text{am}\left(2\sqrt{r}X_0\; \zeta, w\right), \frac{w}{\sqrt{w^2-1}} \right]\right\}\\
+\frac{p X_0}{1-r}\zeta-\frac{1}{2}\,{\rm arccot}\left[\frac{2 \sqrt{r}}{p} \frac{\text{dn}\left(2\sqrt{r}X_0 \;\zeta, w\right)}{\text{sn}\left(2\sqrt{r}X_0 \;\zeta, w\right)}\right]
\end{multline}
and, setting $\zeta=0$ in the above equation, we have
\begin{multline}\label{calphaasafunctionoftheroots}
\alpha = \frac{p X_0}{1-r}+\frac{2p}{r^2-1} \sqrt{\frac{r}{1-w^2}}\\
\times\left\{\Pi\left[\left(\frac{r-1}{r+1}\right)^2,\frac{w}{\sqrt{w^2-1}}\right]\right.\\
\left.-\Pi\left[\left(\frac{r-1}{r+1}\right)^2, \frac{\pi}{2}-\text{am}\left(\sqrt{r}X_0, w\right), \frac{w}{\sqrt{w^2-1}} \right]\right\},
\end{multline}
where we have used that, according to Eq. (\ref{cx0det}), $X_0 \sqrt{r}=2K(\omega)$, and that ${\rm dn}[2K(\omega),\omega]=1$ and ${\rm sn}[2K(\omega),\omega]=0$. The amplitude profile $X_\Phi(\zeta)$ and the angle profile
$\varphi(\zeta)$ are plotted in Fig. 9. Now, using the properties of the ${\rm am}$ and $\Pi$ functions, the above equation can be further simplified to
\begin{multline}\label{calphaasafunctionoftherootsfinal}
\alpha = \frac{p X_0}{1-r}+\frac{4p}{r^2-1} \sqrt{\frac{r}{1-w^2}}\\
\times\Pi\left[\left(\frac{r-1}{r+1}\right)^2,\frac{w}{\sqrt{w^2-1}}\right].
\end{multline}
Recall that the relation of $x_-$ and $x_+$ to $\tau$ and $p$ is given by Eq. (\ref{relxtp}). As in the previous subcase, $X_0$, $x_{-}$ and $x_+$ are known functions of $\tau$ and $p$, and this equation is solved numerically to produce the scaling function for the Casimir force.

From the expressions derived above, it is easy to reproduce the  results previously known for $\alpha=0$. As we will see, this provides a new representation of the older results which is quite convenient for numerical evaluation. First, let us note that, from Eqs. (\ref{alphaasafunctionoftheroots}), (\ref{relxtp}) and (\ref{calphaasafunctionoftheroots}), one immediately obtains $p=0$. Thus, from Eq. (\ref{roots}), it follows that we are in the subcase {\it A)} of real roots. Then $x_+=0, x_-=-(\tau+1)$ for $\tau \ge -1$ and $x_+=-(\tau+1), x_-=0$ for $\tau \le -1$. From Eq. (\ref{x0det}), we find
\begin{equation}\label{x0alphaeq0}
X_0(\tau)=\left\{ \begin{array}{cc}
                    2K\left(\sqrt{-(\tau+1)}\right), & \tau \le -1 \\
                    \frac{2}{\sqrt{\tau+2}} K\left(\sqrt{\frac{\tau+1}{\tau+2}}\right), & \tau \ge -1
                  \end{array}.
\right.
\end{equation}
For the scaling function of the Casimir force, Eq. (\ref{casscalingfunctionpandtau}) gives
\begin{widetext}
\begin{equation}\label{Casimiralphaeq0}
X_{\rm Cas}^{(+,+)}(\tau)=\left\{ \begin{array}{ccc}
                    -4\left(\tau+2\right)^2K^4\left(\sqrt{-(\tau+1)}\right), & \tau \le -1 \\
                    -4 K^4\left(\sqrt{\frac{\tau+1}{\tau+2}}\right), & 0\ge \tau \ge -1\\
                    -4\frac{\tau+1}{(\tau+2)^2} K^4\left(\sqrt{\frac{\tau+1}{\tau+2}}\right),                  & \tau \ge 0
                  \end{array}
\right.
\end{equation}
\end{widetext}
where we have denoted the $\alpha=0$ boundary conditions as $(+,+)$. Denoting the argument of the elliptic $K$ function in a standard way with $k$ and recalling that $x_t=\tau X_0^2$, the above expressions can be rewritten in the form
\begin{widetext}
\begin{equation}
\label{Casimiralphaeq0Krech}
X_{\rm Cas}^{(+,+)}(x_t)=\left\{ \begin{array}{ccc}
                    -4\left(1-k^2\right)^2K^4\left(k\right), & x_t=-4(1+k^2)K^2(k), & x_t \le -\pi^2 \\
                    -4 K^4\left(k\right),&x_t=4(2k^2-1)K^2(k), & 0 \ge x_t \ge -\pi^2\\
                    -4 k^2 (1-k^2) K^4\left(k\right), & x_t=4(2k^2-1)K^2(k), & x_t \ge 0
                  \end{array}.
\right.
\end{equation}
\end{widetext}
The result (\ref{Casimiralphaeq0Krech}) was originally reported in \cite{K97}. The behavior of $X_{\rm Cas}^{(+,+)}(x_t)$ is shown as a thick black line in Fig. \ref{plot:Casimir}.

The scaling function of the Casimir force under $(+,-)$ boundary condition in the Ising mean-field model is
\begin{widetext}
\begin{equation}\label{CasimiralphaeqpiKrech}
X_{\rm Cas}^{(+,-)}(x_t)=\left\{ \begin{array}{ccc}
                    64\; k^2 (1-k^2)\left[K(k)\right]^4, & x_t=-2\left[2K(k)\right]^2(2k^2-1), & x_t \le 0 \\
                    \left[2K(k)\right]^4,& x_t=-2\left[2K(k)\right]^2(2k^2-1), & 0 \le x_t \le 2\pi^2\\
                    \left[2K(k)\right]^4(1-k^2)^2, & x_t=2\left[2K(k)\right]^2(k^2+1), & x_t \ge 2\pi^2
                  \end{array}.
\right.
\end{equation}
\end{widetext}
$X_{\rm Cas}^{(+,-)}(x)$ is plotted (marked with filled circles) in Fig. \ref{plot:Casimir}

Finally, note that the scaling functions $X_{\rm Cas}^{(+,+)}(x)$ and $X_{\rm
Cas}^{(+,-)}(x)$, just derived, are related through \cite{VGMD2009}
\begin{equation}\label{relscfunctionspppm}
X_{\rm Cas}^{(+,+)}(x)=-\frac{1}{4} X_{\rm Cas}^{(+,-)}(-x/2),
\end{equation}
and thus, for the corresponding mean-field Casimir amplitudes, one has
\begin{equation}\label{deltarelpppm}
\frac{\Delta_{\rm Cas}^{(+,+)}}{ \Delta_{\rm Cas}^{(+,-)}}=-\frac{1}{4}.
\end{equation}

\section{Derivation of the low-temperature asymptotic behavior of the Casimir
force within $XY$ Ginzburg-Landau mean-field model under twisted boundary
conditions}
\label{asymptotic}

According to Eq. (\ref{casscalingfunctionpandtau}) when $x_t<0$
\begin{equation}\label{casscalingfunctionpandtauA}
X_{\rm Cas}^{(\alpha)}(\tau)= X_0^4[p^2- \left(1+\tau/2 \right)^2],
\end{equation}
where $\tau$ and $p$ are defined in Eq.(\ref{newvarscaling}). We need to find
the behavior of $X_{\rm Cas}^{(\alpha)}(\tau)$ for $x_t\to -\infty$.

Let us first clarify what is meant by the asymptotic behavior of $\tau$ and $p$ in the
regime $x_t\to-\infty$. For low temperatures one expects
$\Phi(z) \simeq \Phi(z=0)\equiv \Phi_0$ and $d\varphi/dz\simeq \alpha/L$. From
Eq. \eqref{Pvarphidef} and the definition given in Eq.
\eqref{scalingvariables} one then obtains $X_\varphi\simeq\alpha X_0^2$
and, thus, from Eq. \eqref{newvarscaling}, $p\simeq \alpha/X_0$. In terms of
$x_t$ the equation for the order parameter amplitude is given in
\eqref{PhiEqScaled}. Under the assumptions already made, the above equation
becomes $0\simeq\alpha^2 X_0+x_t X_0+2X_0^3$. One concludes that $X_0\gg 1$ with
\begin{equation}\label{X0lowxt}
X_0^2\simeq -(x_t+\alpha^2)/2
\end{equation}
when $x_t \to -\infty$, and that $p^2+\tau+2\simeq 0$, i.e., that $\tau\to -2-p^2$
when $x_t \to -\infty$. Thus, the regime which we need to consider in
(\ref{casscalingfunctionpandtauA}) is $\tau\to -2-p^2$ with $p\simeq
\alpha/X_0\ll 1$. For the Casimir force  from Eq. (\ref{casscalingfunctionpandtauA}) we then obtain
\begin{equation}\label{casscalingfunctionpandtauAs}
X_{\rm Cas}^{(\alpha)}(x_t\to-\infty)\simeq X_0^4[p^2- p^4/4].
\end{equation}

It is easy to check that in the asymptotic regime of interest  $x_\pm$
in Eq. (\ref{roots}) are real. Thus, we need to study the asymptotic behavior
of $X_0$ given by Eq. \eqref{x0det}, taking into account the right-hand side of Eq.
(\ref{alphaasafunctionoftherootsfinal}) which relates $X_0$ to $\alpha$.

Setting
\begin{equation}\label{tauas}
\tau=-2-p^2+a,
\end{equation}
where $p\ll 1$ and $a\to 0$ it is easy to show that Eq. \eqref{x0det} becomes
\begin{equation}\label{x0detAas}
X_0\simeq \ln[16/a],
\end{equation}
while  Eq. (\ref{alphaasafunctionoftherootsfinal}) simplifies to
\begin{equation}\label{alphaasafunctionoftherootsfinalAas}
\alpha= p(X_0-2),
\end{equation}
where we have used Eq. \eqref{x0detAas}. Note that Eqs. \eqref{x0detAas} and \eqref{X0lowxt} imply that $a$ is exponentially small in $\sqrt{|x_t|}$ and, thus, in the remainder we will omit $a$ in Eq. \eqref{tauas} and in any expansion that involves $\tau$. Expressing $p$  from Eq. \eqref{alphaasafunctionoftherootsfinalAas} in terms of $X_0$ and $\alpha$ and inserting the result in Eq. \eqref{casscalingfunctionpandtauAs}, one obtains an expression for the Casimir force in terms of $X_0$ and $\alpha$:
\begin{equation}\label{casscalingfunctionpandtauAsp}
X_{\rm Cas}^{(\alpha)}(x_t\to-\infty)\simeq \alpha ^2 \left(X_0^2+4 X_0+12\right)-\frac{\alpha ^4}{4}.
\end{equation}
Then, making use of  Eq. \eqref{X0lowxt}, one obtains
\begin{equation}\label{casscalingfunctionasympXYA}
X_{\rm Cas}^{(\alpha)}(x_t)\simeq \frac{1}{2}  \alpha^2 \left[|x_t|+4
 \sqrt{2|x_t|}+\frac{1}{2} \left(48-3 \alpha ^2\right)\right],
\end{equation}
where $x_t\to-\infty$. This is the result reported in Eq. \eqref{casscalingfunctionasympXY} in the main text.

\section{Determining the kink temperature in the Ginzburg-Landau model}
\label{tkinkappendix}
We employ the scaled variable defined by Eq. (\ref{scalingvariables}). When the boundary conditions are anti-symmetric, we find a kink in the Casimir force at a temperature $x_{t,\rm kink}$ below the bulk critical temperature. At that point, the two integration constants $X_0$ and $X_\varphi$ both switch between being identically zero ($x_t>x_{t,\rm kink}$) and being positive ($x_t<x_{t,\rm kink}$). Note, however, that the quotient $M=X_\varphi/X_0$ remains non-zero for all temperatures. We determine the kink temperature by enforcing the boundary conditions, Eqs. (\ref{lengthcondition}) and (\ref{twistcondition}).

At the transition point, things are simplified because $X_0,X_\varphi\to0$. The length condition (\ref{lengthcondition}) takes the form
\begin{equation}
\frac{1}{2}=\int_0^\infty \frac{dX_\Phi}{\left(M_{\rm kink}^2+x_{t,\rm kink}X_\Phi^2+X_\Phi^4\right)^{1/2}}. \label{kinkcondition1}
\end{equation}
The twist condition must be treated with more care because the integrand appears to be singular when $X_0=0$. Without taking $X_0$ to zero, (\ref{twistcondition}) may be re-expressed as
\begin{multline}
\frac{\pi}{2}=MX_0\int_{X_0}^\infty \frac{dX_\Phi}{X_\Phi\sqrt{X_\Phi^2-X_0^2}}\\
\times\frac{1}{\left(M^2+x_t X_\Phi^2+X_\Phi^2\left(X_\Phi^2+X_0^2\right)\right)^{1/2}}\\
\equiv f(M,X_0),
\end{multline}
which holds at all temperatures. In particular, just below the kink temperature, $X_0$ is small but non-zero, and (abbreviating $M_k=M_{\rm kink}$ and $x_{t,k}=x_{t,\rm kink}$)
\begin{equation}
\frac{\pi}{2} = f(M_k,0)+\frac{\partial f}{\partial X_0}(M_k,0)\,X_0 + O(X_0^2) \label{ftaylor}
\end{equation}
at that point. If we make the substitution $X_\Phi=X_0 y$, we see that $f(M_k,0)$ is actually non-singular:
\begin{multline}
f(M_k,0)=\left.M_k\int_1^\infty \frac{dy}{y\sqrt{y^2-1}}\right.\\
\left.\times\frac{1}{\left(M_k^2 + x_{t,k} y^2 X_0^2 + y^2\left(y^2+1\right) X_0^4\right)^{1/2}}\right|_{X_0=0}\\
=\int_1^\infty\frac{dy}{y\sqrt{y^2-1}}=\frac{\pi}{2}. \label{fMkzero}
\end{multline}
This means that $(\partial f/\partial X_0)(M_k,0)$ must vanish due to Eq. (\ref{ftaylor}). The derivative is taken most easily from the expression in (\ref{fMkzero}), giving
\begin{multline}
\frac{\partial f}{\partial X_0}(M_k,0)=\left.-M_k X_0\int_1^\infty \frac{y\,dy}{\sqrt{y^2-1}}\right.\\
\left.\times\frac{x_{t,k}+2y^2 X_0^2+2X_0^2}{\left(M_k^2 + x_{t,k} y^2 X_0^2 + y^2\left(y^2+1\right) X_0^4\right)^{3/2}}\right|_{X_0=0}.
\end{multline}
Despite its appearance, this does not trivially vanish when $X_0\to 0$. Instead, restore the original variable $X_\Phi=X_0 y$ to find
\begin{multline}
\frac{\partial f}{\partial\Phi_0}(M_k,0)=\left.-M_k\int_{X_0}^\infty \frac{X_\Phi\,dX_\Phi}{\sqrt{X_\Phi^2-X_0^2}}\right.\\
\left.\times\frac{x_{t,k}+2X_\Phi^2+2X_0^2}{\left(M_k^2+x_{t,k}X_\Phi^2+X_\Phi^2\left(X_\Phi^2+X_0^2\right)\right)^{3/2}}\right|_{X_0=0}
\end{multline}
which suffers no singularity when $X_0$ is replaced by zero. Thus the second condition on $x_{t,k}$ and $M_k$ is
\begin{equation}
0=\int_0^\infty\,dX_\Phi\frac{x_{t,k}+2X_\Phi^2}{\left(M_k^2+x_{t,k}X_\Phi^2+X_\Phi^4\right)^{3/2}}. \label{kinkcondition2}
\end{equation}
Eqs. (\ref{kinkcondition1}) and (\ref{kinkcondition2}) may be recast, with the aid of Eqs. (\ref{cx0det}) and (\ref{calphaasafunctionoftherootsfinal}), into
\begin{equation}
\sqrt{M_k}=2K\left(\sqrt{\frac{1}{2}-\frac{x_{t,k}}{4M_k}}\,\right)
\end{equation}
and
\begin{multline}
\frac{1}{2}M_k\sqrt{x_{t,k}+2M_k}=4M_k K\left(\sqrt{\frac{x_{t,k}-2M_k}{x_{t,k}+2M_k}}\,\right)\\
-(x_{t,k}+2M_k)E\left(\sqrt{\frac{x_{t,k}-2M_k}{x_{t,k}+2M_k}}\,\right),
\label{secondequation}
\end{multline}
which are easily solved numerically to give
\begin{equation}
x_{t,\rm kink}\approx-28.1099,\qquad M_{\rm kink}\approx21.5491.
\end{equation}
In Eq. (\ref{secondequation}), $E(x)$ is the complete elliptic integral of the
second kind.

\end{document}